\documentclass[12pt]{article}
\usepackage{axodraw}
\usepackage{cite}
\usepackage{epsf}
\usepackage{epsfig}

\def\beq{\begin{equation}}
\def\eeq{\end{equation}}
\def\bea{\begin{eqnarray}}
\def\eea{\end{eqnarray}}
\def\barr{\begin{array}}
\def\earr{\end{array}}

\begin{document}

\begin{titlepage}
\begin{flushright}
SHEP-0518, UW/PT 05-16
\end{flushright}
\vskip 0.5cm
\begin{center}
{\Large \bf Finite-Volume Effects for Two-Hadron States
in Moving Frames} \vskip1cm
{\large\bf
C.h.~Kim$^{a}$, C.T.~Sachrajda$^{a}$ and Stephen R.~Sharpe$^{a,b}$ \\
\vspace{.5cm} {\normalsize {\sl $^a$ School of Physics and
Astronomy, Univ. of Southampton,\\ Southampton, SO17 1BJ, UK. \\
\vspace{.2cm} $^b$ Department of Physics, University of
Washington, Seattle, WA-98195-1550, USA .}}}%

\vskip1.0cm


\vskip1.0cm
{\large\bf Abstract\\[10pt]} \parbox[t]{\textwidth}{{
We determine the finite-volume corrections to the spectrum and
matrix elements of two-hadron states in a moving frame, i.e. one
in which the total momentum of the two-hadrons is non-zero. The
analysis is performed entirely within field theory and the results
are accurate up to exponential corrections in the volume. Our
results for the spectrum are equivalent to those of Rummukainen
and Gottlieb which had been obtained using a relativistic quantum
mechanical approach. A technical step in our analysis is a simple
derivation of the summation formulae relating the loop summations
over the momenta of the two hadrons in finite volume to the
corresponding integrals in infinite volume.

}}

\end{center}
\end{titlepage}

\section{Introduction}

Lattice simulations are necessarily performed in a finite volume.
In the evaluation of many quantities of phenomenological interest,
such as hadronic masses or matrix elements with at most a single
hadron in the external states, the errors caused  by the finiteness of
the volume decrease exponentially as the volume is increased~\cite{ml1}.
In practice, as long as the size of the lattice is large enough
compared to the range of the strong interactions, these errors are
numerically negligible. This is not the case, however, when two
(or more) hadrons are present in external states. For such
states the finite-volume effects decrease more slowly, as powers of the
box size, $L$, and need to be understood in order to obtain physical
quantities with good precision.
The theory of such effects has been fully developed for two particles
in their rest frame, i.e. with total momentum $\vec P=0$.
The spectrum of such states was worked out in
refs.~\cite{ml2,ml3,ml4} and the finite volume corrections to
the matrix elements were obtained in refs.~\cite{ll,lmst}.

This paper concerns the extension of these results
to two-particle states with total momentum $\vec P\ne 0$,
which we call a {\em moving frame}. A generalisation of the results
for the spectrum of such states to a moving frame was proposed
some time ago in ref.~\cite{rg}. This was based upon a relativistic
quantum mechanical approach in which the two particles satisfied
a Klein-Gordon equation with an interaction potential of finite range.
Part of the motivation for our project was an attempt to understand
some of the steps in the derivation of ref.~\cite{rg}. We have
therefore revisited the problem using an entirely field theoretic
approach. Although our final result looks superficially quite different
from that obtained in ref.~\cite{rg}, we find that it is possible to
transform our expressions into theirs. Thus we confirm, albeit indirectly,
their assumptions and methodology. Our
derivation and results are useful for a number of reasons. First, the
extension of the results for finite volume corrections to matrix elements
are new. Second, we give an alternative and simple derivation of the
summation formulae that are needed. And, finally, our derivation of
the quantization condition for two-particle states is simple and
transparent, and in particular does not
require the use of an intermediate quantum mechanical theory.

There are several reasons why it is important to have an extension
of the theory of finite volume corrections to a moving frame.
Some were discussed in ref.~\cite{rg}: one obtains additional results
for the scattering phase shifts at different values of
the centre-of-mass momenta from the same lattice configurations;
one can more easily study p-wave decays such as $\rho\to\pi\pi$;
and one avoids the need for vacuum subtractions in s-waves.
An additional important motivation is provided by the need to
control non-perturbative QCD effects in $K\to\pi\pi$ decays, and
hence, for example, to understand quantitatively phenomena such as
the $\Delta I=1/2$ rule and the value of
$\varepsilon^\prime/\varepsilon$. Whilst it is true that the decay
amplitudes can in principle be evaluated in the rest-frame, the
ability to perform simulations in a moving frame will be very
important in controlling systematic uncertainties. For example
chiral perturbation theory is an important guide in performing the
extrapolation of lattice results to those corresponding to
physical $u$ and $d$ quark masses. In this approach there are
a number of unknown constants, and results from moving frames
provide considerably more data points to allow the determination
of these low-energy constants (see, for example, ref.~\cite{spqr}).
Furthermore, as noted above
in the case of scattering amplitudes, the use of a moving frame obviates
the need for vacuum subtractions in the isoscalar channel. This is
important since such subtractions are under poor statistical control.

Throughout this paper we assume that the fields satisfy
periodic boundary conditions in the moving frame (and we
take the temporal extent of the lattice to be infinite). There
has recently been discussion of the possibility of using
anti-periodic~\cite{kimchrist} or twisted
boundary conditions
($\psi(x_i+L)=\exp(i\theta_i)\psi(x_i)\,$ $i=1,2,3$)
in lattice simulations~\cite{twisted}. Although we
do not consider such boundary conditions explicitly in this paper,
the techniques presented below, and in particular the summation
formulae, can readily be generalised for the different cases.

We now briefly present our result for the finite-volume
corrections to the two-particle spectrum in a moving frame. The
example we have in mind is of two pions, and so, for concreteness,
we will call the particles ``pions'' throughout this paper. We
note, however, that the results in this paper can readily be
generalised to the case of two particles with different masses
($m_1$ and $m_2$ say), as long as they are below the inelastic
threshold. An important example is the nucleon-pion system. The
quantization condition given below (see eq.\,(\ref{eq:det}) for
the general case) in terms of the centre-of-mass momentum of each
particle, $q^\ast$, remains valid. $q^\ast$ is then related to the
total centre-of-mass energy $E^\ast$ by the standard kinematical
relation $4q^{\ast\,2}=E^{\ast\,2}
-2(m_1^2+m_2^2)+(m_1^2-m_2^2)^2/E^{\ast\,2}$\,. For the remainder
of this paper we limit the discussion to the case when the two
hadrons are degenerate ($m_1=m_2\equiv m$) but it should be
remembered that the generalisation to non-degenerate particles is
straightforward.

The finite-volume cubic box breaks the Euclidean symmetry and
hence mixes different partial waves. For $\vec P\ne0$, lattice
states which are predominantly s-wave also contain components with
$l=2,4,\dots$ (in contrast to the case with $\vec P=0$ where the
mixing begins at $l=4$). The results presented here were obtained
under the assumption that only scattering in the s-wave is
important in the energy range of interest (which is a good
approximation for $K\to\pi\pi$ decays). The general results are
given in the body of the paper. Let the two-pion state have total
momentum $\vec{P}$ and let $E$ be one of the energy eigenvalues in
the finite-volume $V=L^3$. The total energy in the centre-of-mass
frame is denoted by $E^\ast$ ($E^{\ast\,2}=E^2-P^2$), and $q^\ast$
is defined by $E^{\ast\,2}= 4[q^{\ast\,2}+m^2]$ ($2q^\ast$ is the
magnitude of the relative momentum). We find that the spectrum of
two-pion states is determined by the relation
\begin{equation}
\tan\left[\delta(q^\ast)\right]=-\tan\left[\phi^P(q^\ast)\right]
\,,
\end{equation}
where $\delta(q^\ast)$ is the physical s-wave phase-shift and
the function $\phi^P(q^\ast)$ is defined by
\begin{equation}
\tan\left[\phi^P(q^\ast)\right]=\frac{q^\ast}{4\pi}
\left[c^{P}(q^{\ast\,2})\right]^{-1}\!,
\label{eq:qcfinal0}
\end{equation}
with the box size entering through the following regularized sum
\begin{equation}
c^{P}(q^{\ast\,2})\equiv\frac{1}{L^3}\sum_{\vec{k}}\,\
\frac{\omega_k^\ast}{\omega_k}\
\frac{e^{\alpha(q^{\ast\,2}-k^{\ast\,2})}}{q^{\ast\,2}-k^{\ast\,2}}-
\ {\cal P}\int\frac{d^{\,3}k^\ast}{(2\pi)^3}\
\frac{e^{\alpha(q^{\ast\,2}-\vec{k}^{\ast\,2})}}
{q^{\ast\,2}-\vec{k}^{\ast\,2}}\,.
\label{eq:cpdef0}
\end{equation}
The summation in this expression runs over the usual finite-volume
momenta for periodic boundary conditions:
$\vec{k}=(2\pi/L)\,\vec{n}$, with $\vec{n}$  a vector of integers.
The corresponding pion energy is $\omega_k=\sqrt{\vec{k}^{\,2}+m^2}$.
The quantities $(\omega^\ast_k,\vec{k}^\ast)$
are the energy and momentum of the four-vector
$(\omega_k,\vec{k})$ boosted into the centre-of-mass frame.
${\cal P}$ indicates
that one should take the principal value of the integration and
the superscript {\small $P$} recalls that we are considering a
two-pion system in a moving frame.
The exponential factors in eq.\,(\ref{eq:cpdef0})
regulate the ultraviolet divergences in the sum and integral, and these
divergences cancel in the difference of the two.
The remaining $\alpha$ dependence is exponentially suppressed
as $L\to\infty$. Other choices of ultraviolet regulator are of
course possible, e.g. the analytic regularization used in
refs.~\cite{ml2,ml3,rg}. For this presentation we prefer to
keep a regulator such as the one in (\ref{eq:cpdef0}), so that the
cancellation of UV divergences between the sum and the integral is
manifest.

Our result for the quantization condition,
eq.\,(\ref{eq:qcfinal0}), differs from that
of ref.~\cite{rg} in the form of the kinematical function.
Our $c^P(q^{\ast\,2})$ of eq.\,(\ref{eq:cpdef0}) is replaced in ref.~\cite{rg} by
\begin{equation}\label{eq:crgdef}
\frac{1}{\gamma L^3}\sum_{\vec{k}}\frac{1}{q^{\ast\,2}-r^2}\,
\end{equation}
where $\gamma$ is the Lorentz factor for the
transformation between the moving and centre-of-mass frames,
and~\footnote{%
Ref.~\cite{rg} uses a slightly different definition of
$\vec{r}$ in which the $P/2$ term in eq.~(\ref{eq:rsqdef}) comes
with a positive sign. This difference has no effect on
the sum in eq.~(\ref{eq:crgdef}), however, since a change of
the summation variable $\vec{k} \to \vec{k} + \vec{P}$ brings
our form into that used in Ref.~\cite{rg}. We find the
form in eq.~(\ref{eq:rsqdef}) more natural, as will become
clear in the subsequent discussion.}
\begin{equation}
r^2=\frac{1}{\gamma^2}\left(k_\parallel-\frac{P}{2}\right)^2
+\tilde{k}_\perp^2\,,
\label{eq:rsqdef}
\end{equation}
with $k_\parallel$ and $\tilde{k}_\perp$ the components of $\vec{k}$
parallel and perpendicular to $\vec{P}$, respectively.
The ultraviolet divergences in eq.\,(\ref{eq:crgdef}) are
regulated by analytical regularization.
In sec.~\ref{subsec:comparison} we demonstrate that the expression
for $c^P(q^{\ast\,2})$ in eq.\,(\ref{eq:cpdef0}) is equal to that
in eq.\,(\ref{eq:crgdef}), up to terms which vanish exponentially
with the volume. The two quantization conditions are therefore
equivalent.

We show in sec.~\ref{subsec:comparison} that this equivalence
also extends to the
general situation in which there are multiple partial waves
with non-vanishing phase shifts. In light of this,
the discussion given in ref.~\cite{rg} concerning the
group theoretic constraints on the mixing between partial waves,
and the numerical values of the kinematical functions remains
valid. Thus we do not repeat the discussion in this paper.

In earlier papers $\tan(\phi^P)$ was frequently written in terms
of the dimensionless variable $q^\ast (L/2\pi)$ (and $m (L/2\pi)$
where appropriate). In order to avoid confusion when
comparing our results with earlier ones, we underline that our
notation is different and the energy and momentum variables in
this paper (such as
$\vec{k},\,\omega_k,\,k^\ast,\,\omega_k^\ast,\,q^\ast,\,r$) are
all dimensionful.~\footnote{%
Of course this is a matter of choice.
We choose to use dimensionful variables because it is the physical
energies and momenta which we wish to keep fixed as $L$ is
increased.}

The other principal result of this paper is the expression for the
finite-volume corrections in matrix elements of local
composite operators between external states, one or both of
which contain two hadrons (with an energy below the inelastic
threshold). An important application is to $K\to\pi\pi$ matrix elements.
Our results [see eq.\,(\ref{eq:llmoving}) below] generalise
those of Lellouch and L\"uscher~\cite{ll} and Lin {\em et al.}~\cite{lmst}
for the centre-of-mass frame.

The issues addressed in this paper have also been considered by
Christ, Kim and Yamazaki. Using a related but different approach,
they have also confirmed the validity of the results of
ref.~\cite{rg}, and derived the generalisation of the result of
Lellouch and L\"uscher for moving frames. Their paper is being
released simultaneously with the present article~\cite{cky}.

The remainder of this paper is organized as follows. In the next
section we derive the summation formulae that are needed to
evaluate the generic loop summation over the momenta of
two-particle states. In sec.~\ref{subsec:qcmf} we use the
summation formulae to derive the quantization condition for
two-pion states in finite volume. We then show, in
sec.~\ref{subsec:comparison}, that our results are equivalent to
those obtained in ref.~\cite{rg}, and in sec.~\ref{sec:pert} we
verify the quantization condition at lowest non-trivial order in
perturbation theory in a $\lambda \phi^4$ model of pion
interactions. Section \ref{sec:ll} contains our derivations of the
finite volume corrections to $K\to\pi\pi$ matrix elements. We
present our conclusions in sec.~\ref{sec:conclusions}.

\section{Summation Formulae}\label{sec:sumfor}

Key ingredients in the calculation of finite-volume effects for
two-hadron states are summation formulae relating the sums over
the discrete momenta in a finite volume to the corresponding
integrals over the continuous spectrum in infinite volume.
We start from the Poisson summation formula,
\begin{equation}\label{eq:poisson}
\frac{1}{L^3}\sum_{\vec{k}}\,g(\vec{k}\,)=\int\frac{d^3k}{(2\pi)^3}\,g(\vec{k}\,)
+ \sum_{\vec l\ne\vec{0}}\int\frac{d^3k}{(2\pi)^3} e^{i\, L \,\vec
l\cdot \vec k} g(\vec{k}\,)\,,
\end{equation}
where the summation on the left-hand-side is over
all integer values of $\vec{n}=(n_1,n_2,n_3)$, with
$\vec{k}=(2\pi/L)\,\vec{n}$,
while that on the right-hand-side
is over integer values of $\vec l=(l_1,l_2,l_3)$
excluding $\vec{l}=(0,0,0)$.
In the following, we consider functions
$f(\vec k\,)$ whose Fourier transforms, $\tilde{f}(\vec{r}\,)$,
are non-singular, and are either contained in a finite
spatial region or decrease exponentially as $|\vec{r}\,|\to\infty$.
If we apply the Poisson summation formula to such functions,
the terms with $\vec l\ne 0$ on the right-hand-side of eq.~(\ref{eq:poisson})
decrease at least exponentially as the
box size is sent to infinity, so that
\begin{equation}\label{eq:poisson_f}
\frac{1}{L^3}\sum_{\vec{k}}\,f(\vec{k}\,)=
\int\frac{d^3k}{(2\pi)^3}\,f(\vec{k}\,)
\end{equation}
up to exponentially small corrections.
We note that functions $f(\vec k\,)$ with these properties have
no singularities for real $\vec k$,
and fall off fast enough at $|\vec k|\to\infty$ that the integrals
in eq.~(\ref{eq:poisson}) converge.

We begin our discussion with a review of the summation formulae in
the centre-of-mass frame and then in
section~\ref{subsec:summations_mf} generalise the discussion to
the moving frame.

\subsection{Summation Formulae in the Centre-of-Mass Frame}
\label{subsec:summations_com}

For two-hadron correlators with an energy
below the inelastic threshold, the finite-volume corrections
are contained in summations of the form\footnote{%
Since all variables in this subsection
correspond to the centre-of-mass frame, we do not introduce the
symbol {\small $\ast$} which we use elsewhere in this paper to
distinguish centre-of-mass variables from those in a moving frame.}
\begin{equation}\label{eq:generic_com_sum}
S(\vec{q}\,)\equiv\frac{1}{L^3}\sum_{\vec{k}}\frac{f(\vec{k}\,)}{q^2-k^2}\,
\end{equation}
where we assume that $q^2$ is such
that there is no term in the sum with $k^2\equiv |\vec{k}\,|^2=q^2$ and that
$f(\vec{k}\,)$ has the properties discussed above,
and in particular has no singularities for real $\vec{k}$.
The singularity at $k^2=q^2$ forbids the simple replacement of the
sum with an integral.
Defining spherical coordinates,
$\vec{k}=(k,\theta,\phi)$, we expand $f$ in spherical harmonics
\begin{equation}\label{eq:ylms}
f(\vec{k}\,)=\sum_{l=0}^{\infty}\,\sum_{m=-l}^l\,f_{lm}(k)\,k^l\,
Y_{lm}(\theta,\phi)\,,
\end{equation}
so that (with $q=|\vec{q}\,|$)
\begin{equation}\label{eq:slm}
S(\vec{q}\,)=\sum_{l,m}S_{lm}(q)\qquad\textrm{where}\qquad
S_{lm}(q)\equiv\frac{1}{L^3}\sum_{\vec{k}}\frac{f_{lm}(k)}
{q^2-k^2}\,k^l\,Y_{lm}(\theta,\phi)\,.
\end{equation}
In the following, it is convenient to consider each $S_{lm}(q)$
individually.

To make use of eq.~(\ref{eq:poisson_f}), we subtract from the
summand a function chosen to cancel the pole at $q^2=k^2$, leading
to the result
\begin{eqnarray}
\lefteqn{
\frac{1}{L^3}\sum_{\vec{k}}\frac{f_{lm}(k)-f_{lm}(q)e^{\alpha(q^2-k^2)}}
{q^2-k^2}\,k^l\,Y_{lm}(\theta,\phi)
=} \nonumber \\
&& \int\frac{d^{\,3}k}{(2\pi)^3}\,\frac{f_{lm}(k)-f_{lm}(q)e^{\alpha(q^2-k^2)}}
{q^2-k^2}\,k^l\,Y_{lm}(\theta,\phi)\,,
\label{eq:summationlm}
\end{eqnarray}
which is valid up to terms which are exponentially small in the volume.
The exponential
factors $\exp[\alpha(q^2-k^2)]$ (with $\alpha>0$) are
included so that the subtraction does not introduce ultraviolet
divergences.
Note that the factor of $k^l$ multiplying $Y_{lm}$ is necessary so that
the subtraction does not introduce a singularity at $k=0$,
and thus invalidate the use of eq.~(\ref{eq:poisson_f}).

From eq.\,(\ref{eq:summationlm}) we immediately deduce the required
summation formulae\footnote{%
As above, this equality holds up to exponentially
small corrections, but here, and in similar results below,
this will not be stated explicitly.}
\begin{equation}\label{eq:sflm}
S_{lm}(q)=\delta_{l,0}\ {\cal
P}\int\frac{d^{\,3}k}{(2\pi)^3}\,\frac{f_{00}(k)}{q^2-k^2}\,Y_{00}
+f_{lm}(q){\cal Z}_{lm}(q)\,,
\end{equation}
where
\begin{equation}\label{eq:zlmdef}
{\cal Z}_{lm}(q)=\frac{1}{L^3}\sum_{\vec{k}}
\frac{e^{\alpha(q^2-k^2)}}
{q^2-k^2}\,k^l\,Y_{lm}(\theta,\phi)-\delta_{l,0}\ {\cal
P}\int\frac{d^{\,3}k}{(2\pi)^3}\frac{e^{\alpha(q^2-k^2)}}
{q^2-k^2}\,Y_{00}\,.
\end{equation}
Here we have used the result that the integral on the right-hand side of
eq.\,(\ref{eq:summationlm}) vanishes for $l\ne 0$ as the integrand
is rotationally invariant.
This integral does not vanish, however, for $l=0$, and must be included.
Although the integrand in eq.\,(\ref{eq:summationlm}) has no pole
at $k=q$, in eq.\,(\ref{eq:sflm}) we separate it into two
terms each of which does have such a pole. For consistency, the
two terms need to be regulated in the same way and the principal
value prescription, denoted ${\cal P}$, is a natural choice.

The result in eq.\,(\ref{eq:sflm})
was obtained for $l=0$ in ref.\,\cite{lmst}
[eq.\,(C.24)]
using different methods. Our approach here allows a simple generalisation
to all $l$, and, as will be shown in the next subsection,
a generalisation to a moving frame.
The form of (\ref{eq:sflm}) is intuitively reasonable,
since, when the summand is singular and thus rapidly varying
between grid points, one expects the sum and integral to differ,
with the difference proportional to coefficient of the singularity.
In terms of the Poisson summation formula, eq.~(\ref{eq:poisson}), the
presence of the singularity requires one to keep the terms with $\vec l\ne 0$.

Our result for the kinematical function
${\cal Z}_{lm}(q)$, eq.\,(\ref{eq:zlmdef}),
appears to depend on the parameter $\alpha$.
This, however, is misleading: the $\alpha$ dependence only enters in
the exponentially small terms which we do not control.\footnote{%
Note that for $l=0$ the sum and integral in the definition of ${\cal Z}_{00}$
both have an ultraviolet divergence proportional to $\alpha^{-3/2}$,
but these cancel leaving the exponentially suppressed dependence on $\alpha$.}
This follows from the derivation itself; the key result,
eq.~(\ref{eq:summationlm}), is valid for {\em any} $\alpha$, up to
exponentially suppressed terms. It can also be seen directly by
evaluating the dependence of ${\cal Z}_{lm}$ on $\alpha$, which
has the advantage of showing the explicit form of the $\alpha$
dependence. For example, using the Poisson summation formula
(\ref{eq:poisson}), one can show that
\begin{equation}\label{eq:alphadependence}
\frac{\partial {\cal Z}_{00}}{\partial \alpha}
=
\frac{e^{\alpha q^2} }{(4 \pi \alpha)^{3/2}} \, Y_{00}
\, \sum_{\vec l\ne0}  \,
\exp[ -{\vec l\cdot \vec l\, L^2}/{(4 \alpha)}]
\,.
\end{equation}
The general expression for $\partial {\cal Z}_{lm}/\partial\alpha$
has the same exponentials multiplied by powers of $L$. This
demonstrates that, if $L$ is sent to infinity with $q$ and
$\alpha$ fixed in physical units, the result for ${\cal Z}_{lm}$
is independent of $\alpha$. In practice, when working at fixed
$L$, one should choose $\alpha$ sufficiently small that the
$\alpha$-dependent terms are negligible. The simplest choice is to
send $\alpha\to 0^+$.

This is one example of the more general result that ${\cal Z}_{lm}$ should be
independent of the ultraviolet regulator.
We have checked this numerically by verifying the equality of
our results (in the limit $\alpha\to0^+$) to those
obtained in Refs.~\cite{ml3} using zeta-function
regularization.\footnote{%
The precise relation between our ${\cal Z}_{lm}$ and the $Z_{lm}$
of refs.\,\cite{ml3,rg} is given in eq.~(\ref{eq:cPvsZd}).}

The summation formulae presented above contain those given in
ref.\,\cite{ml2}. The latter are given in
terms of the sum
\begin{equation}\label{eq:generic_com_sumprime}
\frac{1}{L^3}\sum_{\vec{k}}\,\!
^{^\prime}\frac{f(\vec{k}\,)}{q^2-k^2}\,
\end{equation}
with $q^2$ chosen to be one of the values of $k^2$ and the prime
denotes that the terms with $k^2=q^2$ are removed from the sum.
The formulae in ref.\,\cite{ml2} can readily be
obtained by taking the limit $|\vec{q}\,|\to|\vec{K}|$, where
$\vec{K}$ is one of the values of $\vec k$ appearing in the sum in
eq.\,(\ref{eq:generic_com_sum}). This is shown explicitly for the
case $l=m=0$ in appendix C of ref.\,\cite{lmst}.

\subsection{Summation Formulae in a Moving Frame}
\label{subsec:summations_mf}
We now generalise the discussion to a moving frame,
with total energy-momentum $(E,\vec{P}\,)$.
As shown below in sec.\,\ref{subsec:qcmf}, we
require the summation formula for sums of the form
\begin{equation}\label{eq:movingsum}
S(q^\ast)\equiv\frac{1}{L^3}\sum_{\vec{k}}\
\frac{\omega_k^\ast}{\omega_k}\ \frac{f(\vec{k}^\ast)} {q^{\ast\,
2}-k^{\ast\, 2}}\,.
\end{equation}
Note that, while the summation is over
the moving frame momenta $\vec{k}=(2\pi/L)\vec{n}$,
with $\vec{n}$ being a vector of integers, it is convenient to rewrite the
summand in terms of the centre-of-mass momenta $\vec{k}^\ast$
using the Lorentz transformation of eq.(\ref{eq:lorentz}). It
is also convenient, as will be apparent shortly, to pull the Jacobian
$\omega_k^\ast/\omega_k$ out of the function $f(\vec{k}^\ast\,)$.

We now proceed as in sec.\,\ref{subsec:summations_com}, writing
$\vec{k}^\ast$ is terms of spherical polar coordinates,
$\vec{k}^\ast=(k^\ast,\theta^\ast,\phi^\ast)$, and expanding
$f(\vec k^\ast)$ is spherical harmonics
\begin{equation}\label{eq:fkstarylm}
f(\vec{k}^\ast)=\sum_{l=0}^{\infty}\,\sum_{m=-l}^l\,f_{lm}(k^\ast)\,
k^{\ast\,l}\sqrt{4\pi}\,Y_{lm}(\theta^\ast,\phi^\ast)\,.
\end{equation}
The factor of $\sqrt{4\pi}=1/Y_{00}$ is introduced to simplify the subsequent
expressions for $l=0$.
Again using eq.\,(\ref{eq:poisson}) we can write down the required
summation formula
\begin{eqnarray}
\lefteqn{\frac{1}{L^3} \sum_{\vec{k}}\
\frac{\omega_k^\ast}{\omega_k}\ \frac{f_{lm}(k^\ast)-f_{lm}
(q^\ast)e^{\alpha(q^{\ast\,2}-k^{\ast\,2})}}
{q^{\ast\,2}-k^{\ast\,2}}\
k^{\ast\,l}\sqrt{4\pi}\,Y_{lm}(\theta^\ast,\phi^\ast)}
\nonumber\\
&&=\int\frac{d^{\,3}k}{(2\pi)^3}\,\frac{\omega_k^\ast}{\omega_k}\
\frac{f_{lm}(k^\ast)-f_{lm}(q^\ast)
e^{\alpha(q^{\ast\,2}-k^{\ast\,2})}} {q^{\ast\,2}-k^{\ast\,2}}\
k^{\ast\,l}\sqrt{4\pi}\,Y_{lm}(\theta^\ast,\phi^\ast)
\label{eq:intmv}\\
&&=\int\frac{d^{\,3}k^\ast}{(2\pi)^3}\
\frac{f_{lm}(k^\ast)-f_{lm}(q^\ast)e^{\alpha(q^{\ast\,2}-k^{\ast\,2})}}
{q^{\ast\,2}-k^{\ast\,2}}\
k^{\ast\,l}\sqrt{4\pi}\,Y_{lm}(\theta^\ast,\phi^\ast)\,.\label{eq:intcom}
\end{eqnarray}
The Jacobian factor $\omega_k^\ast/\omega_k$ corresponds to the
change of integration variables from the laboratory-frame momenta
$\vec k$ to the centre-of-mass frame momenta $\vec{k}^\ast$. The
integral in eq.\,(\ref{eq:intcom}) is the same as that appearing
in the derivation of the centre-of-mass summation formulae (apart
from the factor of $\sqrt{4\pi}$),
 i.e. the same as the integral
in eq.\,(\ref{eq:summationlm}). It is only non-vanishing for $l=0$.

The summation formulae above can be rewritten into a form
convenient for subsequent manipulations. For $l=0$ we have
\begin{eqnarray}
\frac{1}{L^3}\sum_{\vec{k}}\,\ \frac{\omega_k^\ast}{\omega_k}\
\frac{f_{00}(k^\ast)}{q^{\ast\,2}-k^{\ast\,2}}\
&=&
{\cal P}\int\frac{d^{\,3}k^\ast}{(2\pi)^3}\ \frac{f_{00}(k^\ast)}
{q^{\ast\,2}-k^{\ast\,2}} + f_{00}(q^\ast)\, c^P(q^{\ast\,2})
\,,\label{eq:summationformula}\\
c^P(q^{\ast\,2}) \equiv c^P_{00}(q^{\ast\,2})
&=&
\frac{1}{L^3}\sum_{\vec{k}}\,\
\frac{\omega_k^\ast}{\omega_k}\
\frac{e^{\alpha(q^{\ast\,2}-k^{\ast\,2})}}{q^{\ast\,2}-k^{\ast\,2}}\
- \ {\cal P}\int\frac{d^{\,3}k^\ast}{(2\pi)^3}\
\frac{e^{\alpha(q^{\ast\,2}-k^{\ast\,2})}}
{q^{\ast\,2}-k^{\ast\,2}}
\,,\label{eq:cpdef}
\end{eqnarray}
while for $l>0$ the result is
\begin{equation}
\frac{1}{L^3}\sum_{\vec{k}}\,\ \frac{\omega_k^\ast}{\omega_k}\
\frac{f_{lm}(k^\ast)}{q^{\ast\,2}-k^{\ast\,2}}\
k^{\ast\,l}\sqrt{4\pi}\,Y_{lm}(\theta^\ast,\phi^\ast)
=
f_{lm}(q^\ast)\, c^P_{lm}(q^{\ast\,2})
\,,\label{eq:summlm}
\end{equation}
\begin{equation}
c^P_{lm}(q^{\ast\,2})
=
\frac{1}{L^3}\sum_{\vec{k}}\,\
\frac{\omega_k^\ast}{\omega_k}\
\frac{e^{\alpha(q^{\ast\,2}-k^{\ast\,2})}}{q^{\ast\,2}-k^{\ast\,2}}\
k^{\ast\,l}\,\sqrt{4\pi} \,Y_{lm}(\theta^\ast,\phi^\ast)
\,. \label{eq:cPlmdef}
\end{equation}
Note that we use $c^P$ as a shorthand for $c^P_{00}$. As in the
centre-of-mass frame, the derivation given above shows that the
quantities $c^P_{lm}(q^\ast)$ are independent of $\alpha$ up to
exponentially small corrections in the volume. It is
straightforward to use the expressions above to evaluate the
$c^P_{lm}$ numerically.

It is more difficult than in the centre-of-mass frame to use the
Poisson summation formula to obtain simple analytic expressions
for the $\alpha$ dependence of the $c_{lm}^P$, since the terms
with $\vec l\ne 0$ in eq.~(\ref{eq:poisson}) involve a Fourier
transform with respect to $\vec k$ and not $\vec k^\ast$. We have
studied this issue numerically and have found that the dependence
on $\alpha$ for even $l$ is similar to that found in the
centre-of-mass frame (see eq.\,(\ref{eq:alphadependence})\,). For
odd $l$ on the other hand, the dependence on $\alpha$ appears to
be linear, consistent with the presence of terms proportional to
$\alpha \exp(-m L)$~\footnote{We have checked this in detail for
$\vec P$ lying along an single lattice axis, and for $l\le6$.}.
These differences however, do not affect the essential points
necessary for control of finite volume effects. We repeat that for
fixed $\alpha$ and physical parameters, the $c_{lm}^P$ are
independent of $\alpha$ up to exponentially small corrections in
the volume. In simulations performed at (large) fixed volume,
$\alpha$ should be taken to be sufficiently small so as not to
enhance numerically the terms which are formally exponentially
small in the volume. This can be achieved for example, by taking
the limit $\alpha\to 0+$.

In summary the summation formula for the sum in
eq.\,(\ref{eq:movingsum}) is
\begin{equation}
S(q^\ast)={\cal P}\int\frac{d^{\,3}k^\ast}{(2\pi)^3}\
\frac{f(\vec{k}^\ast)} {q^{\ast\,2}-k^{\ast\,2}}+
\sum_{l=0}^\infty\,\sum_{m=-l}^l\,
f_{lm}(q^\ast)\,c^P_{lm}(q^{\ast\,2})\,.
\end{equation}

\section{Quantization Condition in Moving
Frames}\label{subsec:qcmf}

The two-pion spectrum in finite volume can be determined from the
exponential dependence of correlation functions of composite operators.
We use
\begin{equation}\label{eq:CPdef}
C_{\vec{P}}(t)=
\langle\,0\,|\,\sigma_{\vec{P}}(t)\,\sigma^\dagger(\vec{0},0)\,|\,0\rangle\,
\end{equation}
(with time ordering implicit),
where $\sigma(\vec{x},t)$ is an interpolating operator for
two-pion states and $\sigma_{\vec{P}}(t)$ is its spatial Fourier
transform:
\begin{equation}
\sigma_{\vec{P}}(t)\equiv \int d^3x\,\sigma(\vec{x},t)
\,e^{i\vec{P}\cdot\vec{x}}\,.
\end{equation}
We will consider the two-pion correlation function in energy space
\begin{equation}
\widetilde{C}_{\vec{P}}(E)=\int dt\, e^{-iEt}\,C_{\vec{P}}(t)\,.
\end{equation}
In infinite volume $\widetilde{C}_{\vec{P}}(E)$ has a two-pion cut in the
$s$-plane (where $s=E^2-P^2$ is the Mandlestam variable)
starting at the branch-point at $s=4m^2$.
The quantization of momenta in a finite volume means
that the two-pion cut is replaced by a series of poles and the
energies corresponding to these poles give the two-pion spectrum.
Our aim in this section is to determine the position of these
poles. We restrict the energy $E$ to be in the range $0<s< 16 m^2$, so
that we can neglect finite volume corrections
from four-pion or higher multiplicity intermediate states.
The lower limit on $s$ ensures that the energy is positive in all frames.

Although we phrase our discussion in Minkowski space, we note that
this same object may be obtained from the Euclidean space correlators
calculated in lattice simulations
by analytic continuation to imaginary Euclidean energy
(which is the approach used in ref.\,\cite{ml2}).

The operator $\sigma(\vec x,t)$ must be chosen to have overlap
with whatever lattice two pion states we are interested in,
but the details are not important here. We refer to ref.\,\cite{rg} for
discussion of the symmetry group of two pion
states for different choices of $\vec P$.

For simplicity, and to match the treatment in
refs.~\cite{ml2,rg}, we will take the two pions to be indistinguishable.
Although this restricts the scattering to even $l$, we will not
display this constraint explicitly. This is because the final formulae
can be shown straightforwardly
to hold for all $l$ in the case of distinguishable
particles, although there are differences by factors of 2 at intermediate
stages in the derivation.
We will also assume that there is a symmetry, analogous to G-parity for
physical pions, which forbids intermediate states with odd numbers of pions.
We stress, however, that we make no assumptions about the form of the
pion interaction, {\em e.g.} we do not rely on chiral perturbation theory
or any other expansion scheme. Thus our results are completely general
and apply to any two particle system.

It will be helpful below to use the relations between the
centre-of-mass variables and those in the moving frame. $E$ and
$\vec P$ are the total energy and momentum in the moving frame and
$E^\ast=\sqrt{E^2-P^2}=\sqrt{s}$ is the corresponding total centre-of-mass energy.
The relative velocity ($\vec{\beta}$) between the
centre-of-mass and moving frames is given by
\begin{equation}\label{eq:betadef}
\vec{\beta}=\vec P/E \,, 
\end{equation}
and $E^\ast$ and $E$ are related by
\begin{equation}\label{eq:gammadef}
E=\gamma
E^\ast\quad\textrm{where}\quad\gamma=
1/{\sqrt{\rule{0mm}{3.1mm}1-\beta^2}}\,.
\end{equation}
It is also useful to recall the definition of $q^\ast$, the
magnitude of the relative momentum in the centre-of-mass frame
\begin{equation}\label{eq:qstardef}
q^{\ast\,2}=\frac14 E^{\ast\,2}-m^2\,.
\end{equation}
We stress that, in the following, the
finite volume summations are always over the moving-frame momenta
$\vec k=(2\pi/L)\,\vec{n}\,,$ with $\vec{n}\,$ a vector of integers.

\begin{figure}[t]
\begin{center}
\epsfxsize=\hsize
\epsfbox{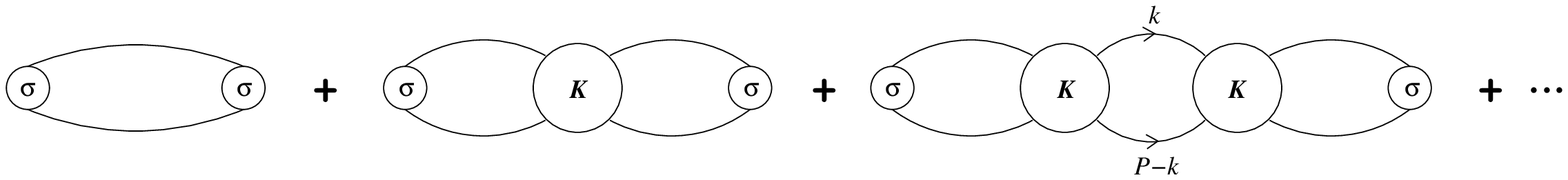}
\end{center}
\caption{Diagrammatic expansion of the correlator $\widetilde C_{\vec P}(E)$.
Propagators are fully dressed and normalized to unity
on shell. $K$ is the amputated two-particle irreducible
four-particle correlation function. The circles at the ends
represent the operator $\sigma$, renormalized by two factors
of $\sqrt{Z}$.}
\label{fig:CP}
\end{figure}

The correlation function $\widetilde C_{\vec P}$ can be expressed in
terms of the Bethe-Salpeter kernel $K$ through the series
shown in fig.\,\ref{fig:CP}. Since we choose $E$ to lie below the
four-pion threshold, there are no intermediate states with four or
more pions and the finite-volume effects in $K$
 are exponentially suppressed~\cite{ml2,lmst}.
The same is true of the dressed single particle propagators~\cite{ml1}.
The only power-law volume corrections arise through the
the two pion loops, and we now turn to an analysis of these corrections.

\subsection{Finite Volume Effects in the Generic Loop
Summation
}\label{subsec:FVloop}

The generic loop integration/summation appearing in fig.~\ref{fig:CP}
is of the form
\begin{equation}
I\equiv\frac{1}{L^3}\sum_{\vec{k}}\,\int
\frac{dk_0}{2\pi}\frac{f(k_0,\vec{k})}{(k^2-m^2+i\varepsilon)
((P-k)^2-m^2+i\varepsilon)}\,
\end{equation}
where $k=(k_0,\vec{k})$ and $P=(E,\vec{P})$ are four vectors and
we have left out a factor of $i^2=-1$ from the propagators which
will be accounted for later. The function $f$ contains the
energy-momentum dependence arising from the kernels on either
``side'' of the loop (with one of these kernels replaced by the
matrix element of the operator $\sigma$ if the loop lies at one of
the ``ends'' of bubble chains shown in fig.~\ref{fig:CP}) as well
as that from the dressed propagators. Since $E$ and $\vec P$ are
fixed, $f$ can be expressed as a function of the four-vector
$(k_0, \vec k)$ alone. Note that the dressing of the propagators
does not effect the allowed values of the three momenta, which are
constrained as usual by the finite volume. The only properties of
$f$ that we need are, first, that it has no singularities for real
$\vec k$ (which holds given our kinematical constraint on $E$),
and, second, that its ultraviolet behaviour is such as to render
the integral and sum convergent.

The power-law volume corrections arise, as usual, because the integrand/summand is
singular. To simplify the pole structure we first perform the $k_0$ integration.
We choose to close the contour of integration so as to
pick up the ``particle'' contribution from the first pole and
the ``anti-particle'' contribution from the second:
\begin{equation}
I=-i\frac{1}{L^3}\sum_{\vec{k}}\left\{\frac{f(\omega_k,\vec{k})}{2\omega_k
((E-\omega_k)^2-\omega_{Pk}^2)}+
\frac{f(E+\omega_{Pk},\vec{k})}{2\omega_{Pk}
((E+\omega_{Pk})^2-\omega_k^2)}
\right\}\,,\label{eq:iresult}\end{equation} where
\begin{equation}
\omega_k=\sqrt{\vec{k}^2+m^2}\quad\textrm{and}\quad
\omega_{Pk}=\sqrt{(\vec{P}-\vec{k})^2+m^2}\,.
\end{equation}
We have dropped the factors of $i\varepsilon$ since what remains
is a summation, and there is no obstruction to setting $\varepsilon=0$, as long
as the poles do not coincide with the allowed values of $\vec k$. This will be
true in general, since the energies of interest are shifted by interactions away
from those of two free particles. This argument does not hold in infinite volume,
where we must perform an integral over $\vec k$, and the
factors of $i\varepsilon$ must be kept. We return to this
point below.

For the kinematic region of interest, $0<E^2-P^2<16 m^2$, it is
straightforward to show that the only singularity in $I$ is
the explicit pole in the first term inside the braces in
eq.\,(\ref{eq:iresult}), which occurs at those values of $E$ for which there is
a term in the summation with $\omega_k+\omega_{Pk}=E$.
It is this
singularity which leads to finite-volume corrections which
decrease like powers of the volume. There are no other poles
(either explicit or hidden in $f$) for
$E$ in this kinematical range. For the second term in
eq.\,(\ref{eq:iresult}) we can therefore replace the summation by
the corresponding integration and write
\begin{equation}
I=I_1+I_2\,,
\end{equation}
where
\begin{eqnarray}
I_1&=&-i\frac{1}{L^3}\sum_{\vec{k}}\frac{f(\omega_k,\vec{k})}{2\omega_k
((E-\omega_k)^2-\omega_{Pk}^2)}\qquad\textrm{and}\\
I_2&=&  -i\int \frac{d^3 k}{(2\pi)^3}\,
\frac{f(E+\omega_{Pk},\vec{k})}{2\omega_{Pk}
((E+\omega_{Pk})^2-\omega_k^2)}\,.
\end{eqnarray}

To determine the finite-volume corrections we need to
examine $I_1$ in more detail. It is convenient to change variables
in the summand to those in the centre-of-mass frame. In the moving
frame $\vec k=(k_\parallel,\tilde k_\perp)$ is the momentum of an
on-shell particle with mass $m$, where $k_\parallel$ and $\tilde k_\perp$
are the components parallel and perpendicular to $\vec\beta$
respectively. The corresponding momentum and energy in the
centre-of-mass frame, $(k^\ast_\parallel,\tilde{k}^\ast_\perp)$
and $\omega_k^\ast$, are given by
\begin{equation}\label{eq:lorentz}
k^\ast_\parallel=\gamma(k_\parallel-\beta\omega_k),\quad
\tilde{k}^\ast_\perp=\tilde{k}_\perp\quad\textrm{and}\quad
\omega_k^\ast=\sqrt{k^{\ast\,2}+m^2}=\gamma(\omega_k-\beta
k_\parallel)\,.
\end{equation}
We now rewrite $I_1$ in the form
\begin{equation}
I_1=-i\frac{1}{L^3}\frac{1}{E^\ast}\sum_{\vec{k}}\frac{1}{2\omega_k}
\frac{f^\ast(\vec{k}^\ast)}
{E^\ast-2\omega_k^\ast}=\,
-i\frac{1}{L^3}\frac{1}{2E^\ast}\sum_{\vec{k}}\frac{\omega_k^\ast}{\omega_k}
\,\frac{f^\ast(\vec{k}^\ast)}{q^{\ast\,2}-k^{\ast\,2}}\,\frac
{E^\ast+2\omega_k^\ast}{4\omega_k^\ast}\,,
\end{equation}
where $f^\ast$ is the function $f$ rewritten in terms of the
centre-of-mass variables. Since $\omega_k^\ast$ is a dependent variable,
we choose to write $f^\ast$ as a function of $\vec{k}^\ast$ alone.

As expected, we see that the poles
correspond to those terms for which $\omega_k^\ast=E^\ast/2$ or
equivalently $k^{\ast\,2}=q^{\ast\,2}$\,.
We have deliberately not rewritten
the factor $1/\omega_k$ in terms of the centre-of mass variables,
so that the summand contains the same Jacobian as in
the summation formulae (\ref{eq:summationformula}) and
(\ref{eq:summlm}). Using these formulae, we find
\begin{equation}
I_1=-i\frac{1}{2E^\ast}\,{\cal P}\int\frac{d^3k^\ast}{(2\pi)^3}
\,\frac{f^\ast(\vec{k}^\ast)}{q^{\ast\,2}-k^{\ast\,2}}\,\frac
{E^\ast+2\omega_k^\ast}{4\omega_k^\ast}\,-\,\frac{i}{2E^\ast}
\sum_{l=0}^{\infty}\sum_{m=-l}^{l}
 f^\ast_{lm}(q^\ast)\,c^{P}_{lm}(q^{\ast\,2})\,,
\label{eq:i1v1}\end{equation}
where we recall that $f^\ast_{lm}$ is defined as in
eq.\,(\ref{eq:fkstarylm}) and $c^{P}_{lm}(q^{\ast\,2})$ are
defined in eqs.\,(\ref{eq:cpdef}) and (\ref{eq:cPlmdef}).

There is one further step we need to perform in order to write
$I_1$ as the infinite-volume result together with a correction,
i.e. we should replace the principal-value integral in
eq.\,(\ref{eq:i1v1}) by the corresponding integral with the
Feynman $i\varepsilon$ prescription in the propagator and
a ``delta-function'' term:
\begin{eqnarray}
I_1&=&-i\frac{1}{2E^\ast}\,\int\frac{d^3k^\ast}{(2\pi)^3}
\,\frac{f^\ast(\vec{k}^\ast)}{q^{\ast\,2}-k^{\ast\,2}+i\varepsilon}\,\frac
{E^\ast+2\omega_k^\ast}{4\omega_k^\ast}
\nonumber\\
&&\mbox{}\qquad
+\frac{q^\ast\,f^\ast_{00}(q^\ast)}{8\pi E^\ast} -\,
\frac{i}{2E^\ast}
\sum_{l=0}^{\infty}\sum_{m=-l}^{l}
 f^\ast_{lm}(q^\ast)\,c^{P}_{lm}(q^{\ast\,2})
\,.
\label{eq:i1v2}\end{eqnarray}
Note that the ``delta-function'' term picks out the $l=0$ part of $f^\ast$.
Observing that the first term in eq.\,(\ref{eq:i1v2}) is
exactly the infinite volume expression for $I_1$ in Minkowski space
(after retracing the steps in the derivation above),
we arrive at our final result,
\begin{eqnarray}
I&=&I_\infty+ I_{FV}
\label{eq:ifinal}\\
I_{FV} &=& \left\{\frac{q^\ast\,f^\ast_{00}(q^\ast)}{8\pi
E^\ast} -\,\frac{i}{2E^\ast}\,\sum_{l=0}^{\infty}\sum_{m=-l}^{l}
f^\ast_{lm}(q^\ast)\, c^{P}_{lm}(q^{\ast\,2})\right\}\,,
\label{eq:IFV}
\end{eqnarray}
where $I_\infty$ is the
infinite volume result for the original loop integral,
$I_\infty=I_{1,\,\infty}+I_2$.

The quantity $I_{FV}$ is the desired finite volume correction.
Note that it involves the function $f^\ast$ with momentum
argument $k^\ast=q^\ast$, so that both particles are on mass shell.
Thus the infinite volume amplitudes contained in $f^\ast$ are
evaluated with physical momenta, just as if one were
picking out the cut in an infinite volume correlator.
What is striking about this result is that $q^\ast$ is a momentum
determined by the choice of $E$, and can take any positive value.
In particular, $q^\ast$ is not constrained by
the finiteness of the volume: it does not have to take
on one of the (boosted) free particle values.
This is a crucial result in the next section where it
allows the finite volume corrections to be expressed in
terms of physical scattering amplitudes, with the finite
volume entering through the purely kinematic functions $c^P_{lm}$.

\subsection{Quantization Condition}\label{subsec:qc}

In this section we use the result in eqs.\,(\ref{eq:ifinal}) and
(\ref{eq:IFV}) to derive
an expression for the two-pion energies in finite volume.
We first derive the general, though formal, result, next work out the
simplest case in which only the s-wave scattering amplitude
is non-vanishing, and finally give the result assuming that
the scattering amplitude vanishes for $l>l_{\rm max}$.

\begin{figure}[t]
\begin{center}
\epsfxsize=\hsize
\epsfbox{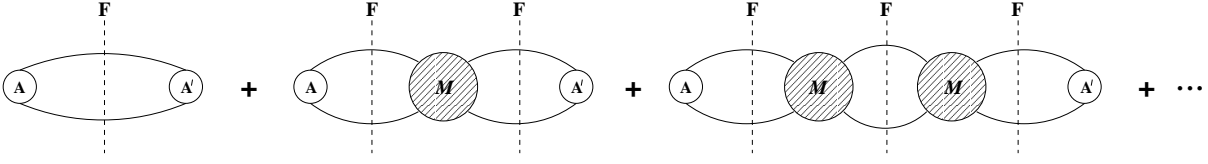}
\end{center}
\caption{Contributions to the volume dependent part
of the $\sigma$ correlator, $\widetilde C_{\vec P}^{FV}$.
The notation is as in fig.~\protect\ref{fig:CP} except that the
filled circles represent the full scattering amplitude, $M$,
given by a geometric sum of any number of
insertions of the kernel $K$, and the vertical dashed lines
indicate that the on-shell finite volume part, $I_{FV}$, has
been used for the loop integral. The quantities $A$ and $A'$ are
defined in the text.}
\label{fig:CPFV}
\end{figure}

Our approach is straightforward: we use
eq.\,(\ref{eq:ifinal}) for each of the loop summations in
fig.~\,\ref{fig:CP}, and then reorganize the series.
If we keep $I_\infty$ in each loop then we recover
the infinite volume correlator $\widetilde{C}_{\vec{P}}^\infty(E)$.
As already discussed, it does
not contain the poles which we are seeking. The part of
interest is the finite volume correction
\begin{equation}
\widetilde{C}_{\vec P}^{FV}(E) =
\widetilde{C}_{\vec P}(E)-\widetilde{C}_{\vec P}^\infty(E)
\,.\label{eq:CPFV}
\end{equation}
This contains the desired poles, as well as cuts which
must cancel those in $\widetilde{C}_{\vec P}^\infty(E)$.
Diagrammatically, $\widetilde{C}_{\vec P}^{FV}$ is obtained by keeping
at least one insertion of $I_{FV}$. These contributions are shown
in fig.\,\ref{fig:CPFV}, and lead to the following general result:
\begin{eqnarray}
\widetilde{C}_{\vec P}^{FV}(E) &=& -A'\; F\; A + A'\;F\;(iM/2)\;F \; A + \dots
\label{eq:CPFVresa}\\
&=& - A'\; F\; \frac{1}{1 + i M F/2}\; A
\,.\label{eq:CPFVres}
\end{eqnarray}
Here we have taken into account the factor of
$i^2$ dropped from the loop in the previous section, as well as
symmetry factors
of $1/2$ arising from the identical nature of the particles.

The quantities in eqs.\,(\ref{eq:CPFVresa}) and (\ref{eq:CPFVres})
 act on the vector space of states
 of two pions with total energy $E$ and momentum $\vec{P}$.
In particular, $A$ and $A'$  are given by
\begin{equation}
A = {}_{\rm out}\langle \pi \pi\;;E,\vec P |\sigma^\dagger(0,\vec 0)
| 0 \rangle  Z_\pi \,,\qquad
A' = Z_\pi \langle 0 | \sigma(0,\vec 0)| \pi \pi\;;E,\vec P\rangle_{\rm in}
\,,
\end{equation}
and involve a geometric series of contributions with any number of
insertions of the kernel $K$, with the intermediate loops containing
the infinite-volume expression $I_\infty$. The factor of
$i\varepsilon$ in $I_\infty$ leads to the presence of in- and out-states as shown.
$M$ is the on-shell scattering amplitude, obtained by summing
a similar geometric series. And, finally, $F$ is a kinematic
factor determined by the form of $I_{FV}$, which, in particular,
accounts for the fact that the ``cut'' can have non-vanishing angular
momentum.

To give these quantities a concrete representation we
change to centre-of-mass variables, since we can then
use the partial wave basis. In particular, $M$ is diagonal in
this basis, with matrix elements
\begin{equation}
M_{l_1,m_1;\,l_2,m_2} = \delta_{l_1 l_2}\; \delta_{m_1 m_2}
\frac{16\pi E^\ast}{q^\ast}\,
\frac{\left(\exp[{2i\delta_{l_1}(q^\ast)}]-1\right)}{2 i}
\,,\label{eq:Mlm}
\end{equation}
where $\delta_l$ is the phase shift in the $l$-th partial wave.
For $I_{FV}$ we have already changed to centre-of-mass variables in
the result (\ref{eq:IFV}), from which it follows, using
the definition of $f^\ast_{lm}$ [see eq.\,(\ref{eq:fkstarylm})],
and the completeness of the spherical harmonics,
that
\begin{eqnarray}
F_{l_1,m_1;\,l_2,m_2} &=& \frac{q^\ast}{8\pi E^\ast}
\left( \delta_{l_1 l_2} \delta_{m_1 m_2}
+ i F^{FV}_{l_1,m_1;\,l_2,m_2} \right)
\label{eq:Fgenres}\\
F^{FV}_{l_1,m_1;\,l_2,m_2} &=& -\frac{4\pi}{q^\ast}\,
\sum_{l=0}^{\infty} \sum_{m=-l}^{l}
\frac{\sqrt{4\pi}}{q^{\ast\,l}} c^{P}_{lm}(q^{\ast\,2})
\int d\Omega^\ast \, Y^\ast_{l_1,m_1}
\,Y^\ast_{l,m}\, Y_{l_2,m_2}
\,, \label{eq:FFV}
\end{eqnarray}
where the implicit arguments of the spherical harmonics
are $(\theta^\ast,\phi^\ast)$, and the asterixes on the spherical
harmonics indicate complex conjugation.
The integral can be expressed in terms of Wigner 3-j symbols, but an
explicit expression will not be needed.

As for the unknown matrix elements
$A$ and $A'$, the change of variables leads to no simplification,
but no complication either. They are simply unknown
vectors $A_{lm}$ and $A'_{lm}$. Note that, unlike $M$, the matrix
elements $A$ and $A'$ need not be Lorentz invariant,
since the operator $\sigma$ can involve spatially separated
pion fields. But this lack of invariance is of no consequence for
our argument, since the role of $A$ and $A'$ is simply to provide
the coupling of the external operators to the on-shell two pion
states. In other words, we are interested here in the
positions of the poles, and not their residues.

The positions of the poles in $\widetilde{C}_{\vec P}^{FV}$
are determined by the factor which lies between
$A'$ and $A$, namely $F (1 + i M F/2)^{-1}$. $F$ itself inherits from
$c^P_{lm}$ poles at all energies corresponding to two free pions both having
allowed momenta in the moving frame, as is clear from eqs.\,(\ref{eq:cpdef},\
\ref{eq:cPlmdef}). Turning on the interaction will, however, shift these
poles, and indeed one can see that they cancel between numerator and
denominator if $M\ne 0$. The poles are shifted to the positions
determined by requirement that $1 + i M F/2$ have zero eigenvalue,
or equivalently by the formal quantization condition:
\begin{equation}
\mathrm{det}(1 + i M F/2) = 0
\,.\label{eq:det}
\end{equation}
This is the main result of this section. It demonstrates that,
for any $\vec P$, the finite volume energy shifts
depend on interactions only through the infinite volume
scattering amplitude,
as was shown by L\"uscher in ref.\,\cite{ml3} for $\vec P=0$.
The result (\ref{eq:det}) also gives a clear separation between
the dynamical effects (the scattering amplitude $M$, independent of
the volume) and the kinematical effects of finite volume
(the matrix $F$, dependent on the volume, but independent of the
scattering amplitude).

In the derivation of ref.\,\cite{ml3} it was important that
the range of the effective non-relativistic
potential, $R$, satisfy $L/2 > R$, aside from
an exponentially falling tail. Without this condition there
is no ``outside'' region where the wavefunction is composed
of free waves with phase determined by the phase shift. Thus
for $L/2 \le R$ one would not expect the energy shift to
depend only on the phase shifts, i.e. on $M$ alone.
By contrast, our result, eq.\,(\ref{eq:det}),
appears to imply that the energy shifts depend only on $M$
without any proviso. This is, however, misleading. Our entire
approach
relies on the assumption that sums of non-singular functions
in momentum space can be well approximated by integrals. When the
range of the potential is larger than the box size, the spread
of its Fourier transform ($\Delta k \sim 1/L$)
is smaller than the spacing between the discrete finite volume
momenta. Consequently, the starting assumption of our derivation fails,
and there is no conflict with the result of ref.\,\cite{ml3}.

To make the result (\ref{eq:det}) into a practical tool we need,
following refs.\,\cite{ml3,rg,lmst},
to truncate the number of partial waves that contribute.
In other words, we assume that $\delta_l=0$ for $l> l_{\rm max}$.
In the following subsection we discuss the simplest case, $l_{\rm max}=0$,
and then return to the general case where $0 < l_{\rm max} < \infty$.

\subsubsection{Quantization Condition: s-wave Interactions}
\label{subsec:qcswave}

We assume that $A$ and $A'$ couple the external operators to
two pions with an s-wave component, and that only the s-wave part of
the scattering amplitude is non-vanishing. Note that although
$F$ couples the s-wave to other waves [through the
$c^P_{lm}$ terms with $l\ne 0$ in $F^{FV}$, eq.\,(\ref{eq:FFV})],
these couplings would have to contribute through terms of the form $F\,M\,F$,
$F\,M\,F\,M\,F$,  etc. in eq.\,(\ref{eq:CPFVresa})
but these vanish since $M=0$ for the higher waves.
Thus the geometric series leading
to the pole occurs only in the s-wave, and can be treated
in isolation.
We discuss this point more thoroughly in the general context
of the next subsection.

Since the problem is now one-dimensional, the quantization condition is
simply
\begin{equation}
1 + i M_s F_s/2 = 0
\,,
\label{eq:swavecondition}
\end{equation}
where
\begin{equation}
M_s =\frac{16\pi E^\ast}{q^\ast}\,
\frac{\left(e^{2i\delta(q^\ast)}-1\right)}{2 i}
\,,
\label{eq:mtodelta}
\end{equation}
with $\delta$ the s-wave phase shift,
and $F_s$ is the $l=0$ component of $F$:
\begin{equation}
F_s \equiv F_{00;\,00} = \frac{q^\ast}{8\pi E^\ast}
-\,\frac{i}{2E^\ast}\,c^{P}(q^{\ast\,2})\,.
\end{equation}
We note that the quantization condition (\ref{eq:swavecondition})
has been obtained in ref.~\cite{beaneetal} using a similar
method to ours in the context of a non-relativistic effective
theory of nucleon-nucleon scattering.
After rearrangement, the quantization condition can
be written in the form
\begin{equation}
\tan\left[\delta(q^\ast)\right] = -\tan\left[\phi^P(q^\ast)\right]
\quad\textrm{where}\quad
\tan\left[\phi^P(q^\ast)\right]=\frac{q^\ast}{4\pi}
\left[c^{P}(q^{\ast\,2})\right]^{-1}\!.
\label{eq:qcfinal}
\end{equation}
 Thus the finite-volume two-pion
states on a cubic lattice of size $L^3$ are those with energies
(and hence values of $q^\ast$) which satisfy
eq.\,(\ref{eq:qcfinal})\,. For the particular case of $\vec{P}=0$
one immediately recovers the L\"uscher condition\,\cite{ml3}\,.

We note in passing that for external operators in the
moving frame having no centre-of-mass s-wave component,
the allowed finite volume energies are those of two free particles.

\subsubsection{Quantization Condition: general interactions}
\label{subsec:qcgeneral}

In this subsection we assume that the phase shifts, and
consequently scattering amplitudes, are small enough that
they can be ignored for $l > l_{\rm max}$.
In other words, we assume that
\begin{equation}
M_{l,m;\,l,m} = 0 \quad \textrm{if}\quad l > l_{\rm max}\,.
\label{eq:truncatedM}
\end{equation}
As we now explain, this turns out to imply that one can
truncate the entire problem to the subspace with $l \le l_{\rm max}$.
In particular the determinant of the full matrix
$(1 + i M F/2)_{l_1,m_1;\,l_2,m_2}$ vanishes if the determinant of the block
with $l_1,l_2\le l_{\rm max}$ vanishes. Thus the
quantization condition becomes
\begin{equation}
\mathrm{det}(1 + i M F/2)\bigg|_{l_1,l_2\le l_{\rm max}} = 0
\,.
\label{eq:truncateddet}
\end{equation}
The condition (\ref{eq:swavecondition})
of the previous section is the simplest
example ($l_{\rm max}=0$) of this general result.

The subtlety in establishing this result arises
from the fact that $F$ has
non-vanishing elements connecting $l_1 \le l_{\rm max}$
and $l_2 > l_{\rm max}$ and vice-versa. To see that these
``block-off-diagonal'' matrix elements are irrelevant, it is
useful to introduce a projection operator, $P$, onto the
subspace with $l\le l_{\rm max}$. Then the assumed form
of the scattering amplitude, eq.~(\ref{eq:truncatedM}),
can be written as $M = P M P$. Using this property,
and denoting the truncation of $F$ to the $l\le l_{\rm max}$
block by $F_P \equiv PFP$, one can manipulate the
result for the finite-volume part of the correlation function
into the form
\begin{equation}
\widetilde{C}_{\vec P}^{FV}(E) =
-A'\, \left[
F - (i/2) F M \frac{1}{P + i M F_P/2} F
\right]\, A
\,.
\end{equation}
This shows that poles occur when the truncated
matrix $P+i M F_P/2$ has zero eigenvalues, which
is equivalent to the condition (\ref{eq:truncateddet}).

A consequence of this argument is that we only need to determine
$F$ in the truncated basis. Using the explicit form in
eqs.(\ref{eq:Fgenres}) and (\ref{eq:FFV}), we see that if $l_1,l_2
\le l_{\rm max}$, then the angular integral in $F^{FV}$ vanishes
for $l>l_1+l_2$. Thus only a finite number of the quantities
$c^P_{lm}$ are needed to calculate the truncated determinant.

In order to compare our result to that of refs.~\cite{ml3,rg}, it
is convenient to pull out the volume-independent first term from
$F$ in eq.\,(\ref{eq:Fgenres}). Then one can rewrite the quantization
condition as
\begin{equation}
\mathrm{det}(\cos\delta - \sin\delta\,
F^{FV})\bigg|_{l_1,l_2\le l_{\rm max}} = 0
\,, \label{eq:detRGform}
\end{equation}
where $\cos\delta$ is a diagonal matrix with entries
$\cos\delta_{l_1}$ on the diagonal, and similarly for $\sin\delta$.
This condition has the same form as eq.\,(100) of ref.~\cite{rg},
and agrees if our matrix $F^{FV}$ equals that denoted ${\cal M}$ there
[and by ${\cal M}^{RG}$ in eq.~(\ref{eq:FFVvsMRG}) below].
We have checked, using the relation between our $c^P_{lm}$ and the
zeta-functions $Z^d_{lm}$ of ref.~\cite{rg} given in eq.\,(\ref{eq:cPvsZd})
below, that in fact,
\begin{equation}
F^{FV}_{l_1,m_1;\,l_2,m_2} =
i^{-l_1}\ {\cal M}^{RG}_{l_1,m_1;\,l_2,m_2} \ i^{l_2}
\,.\label{eq:FFVvsMRG}
\end{equation}
The extra factors of $i$ cancel in the determinant in
eq.~(\ref{eq:detRGform}), because $\cos\delta$ and $\sin\delta$
are diagonal matrices.
Thus our quantization condition agrees with that of ref.~\cite{rg}.

\section{Comparison of Quantization Condition with that of
Rummukainen and Gottlieb}\label{subsec:comparison}

In this section we show that the kinematical functions
$c^P_{lm}(q^{\ast\,2})$, defined in eqs.\,(\ref{eq:cpdef}) and
(\ref{eq:cPlmdef}), can be rewritten in the form given in
ref.~\cite{rg}. As already noted in the previous section, this
implies the equivalence of the quantization conditions. To compare
the results requires a small kinematical exercise, i.e. the
rewriting of the factor
$(\omega_k^\ast/\omega_k)/(q^{\ast\,^2}-k^{\ast\,^2})$, which
appears in the definition of $c^P_{lm}(q^{\ast\,2})$ in terms of
the momentum $r$ of eq.\,(\ref{eq:rsqdef}). We need this in the
vicinity of the pole, i.e. for $q^{\ast\,2}\approx k^{\ast\,2}$.
We first recall that, for a given choice of $\vec k$, setting
$q^{\ast\,2}=k^{\ast\,2}$ corresponds to a kinematical situation
with both pions on-shell. This follows from the derivation in the
previous section, in which the pole in $q^\ast$ arises when both
intermediate propagators are on shell. The particular consequence
we need here is that $\omega_k^\ast=E^\ast/2$ at the pole, since
the two on-shell particles share the energy equally in the CM
frame. This result can be seen directly from
\begin{equation}
\omega_k^{\ast\,2}-(E^\ast/2)^2 \equiv(k^{\ast\,2}+m^2) - (q^{\ast\,2}+m^2)
= k^{\ast\,2} - q^{\ast\,2}
\,.\label{eq:onshellres}
\end{equation}
An important corollary is that $\vec r=\vec k^{\ast}$ at the pole.
To see this we recall from ref.\,\cite{rg} that
\begin{equation}
r_\parallel\equiv \gamma^{-1}(k_\parallel-P/2)\,,\quad
\tilde r_\perp \equiv \tilde k_\perp = \tilde k^\ast_\perp\,.
\label{eq:rdef}
\end{equation}
Using the inverse Lorentz transform,
$k_\parallel=\gamma(k^\ast_\parallel+\beta \omega_k^\ast)$,
together with $E^\ast=E/\gamma$ and $\beta=P/E$,
one finds
\begin{equation}
r_\parallel=k^\ast_\parallel + \beta \omega_k^\ast - P/(2\gamma)
= k_\parallel^\ast + \beta(\omega_k^\ast-E^\ast/2)
\,.\label{eq:kstvsr}
\end{equation}
Since $\omega^\ast_k=E^\ast/2$ at the pole, this establishes the
desired result. Thus setting $q^{\ast\,2}=k^{\ast\,2}$ implies
also $q^{\ast\,2}=r^2$, and it follows that
the positions of the poles in $q^\ast$
in the sum contained in $c^P$, eq.\,(\ref{eq:cpdef}),
are the same as those in the sum of ref.\,\cite{rg},
eq.\,(\ref{eq:crgdef}). This is clearly a necessary condition for
the equivalence of the results.

It is now straightforward to determine the residue of the pole
when written in terms of $r^2$ rather than $k^{\ast\,2}$.
Working to linear order in the difference $E^\ast/2-\omega_k^\ast$
we have
\begin{equation}
E^\ast/2-\omega_k^\ast \approx
 \frac{E^{\ast\,2}/4-\omega_k^{\ast\,2}}{2\omega_k^\ast}
= \frac{q^{\ast\,2}-k^{\ast\,2}}{2 \omega_k^\ast}
\,.
\end{equation}
 Using this result in eq.\,(\ref{eq:kstvsr}) yields
\begin{equation}
k^{\ast\,2} - r^2
\approx
2 \beta k^\ast_\parallel (E^\ast/2-\omega_k^\ast)
\approx
\beta k^\ast_\parallel (q^{\ast\,2}-k^{\ast\,2})/\omega_k^\ast
\,.
\end{equation}
Thus we find
\begin{eqnarray}
q^{\ast\,2}-r^2 &=& (q^{\ast\,2}-k^{\ast\,2})
(1 + \beta k^\ast_\parallel/\omega_k^\ast)
+ O\left[(q^{\ast\,2}-k^{\ast\,2})^2\right] \\
&=&
(q^{\ast\,2}-k^{\ast\,2})
(\omega_k/\gamma\omega_k^\ast) +
 O\left[(q^{\ast\,2}-k^{\ast\,2})^2\right]
\,, \label{eq:residuerelation}
\end{eqnarray}
where in the second step we have used
$\omega_k=\gamma(\omega_k^\ast + \beta k^\ast_\parallel)$.

Using this result we can rewrite our kinematical functions in
terms of $r^2$. We begin with $c^P(q^{\ast\,2})$:
\begin{eqnarray}
c^{P}(q^{\ast\,2})&=&\frac{1}{L^3}\sum_{\vec{k}}\,\
\frac{\omega_k^\ast}{\omega_k}\
\frac{e^{\alpha(q^{\ast\,2}-k^{\ast\,2})}}{q^{\ast\,2}-k^{\ast\,2}}- \
{\cal P}\int\frac{d^{\,3}k}{(2\pi)^3}\
\frac{\omega_k^\ast}{\omega_k}\
\frac{e^{\alpha(q^{\ast\,2}-k^{\ast\,2})}} {q^{\ast\,2}-k^{\ast\,2}}\\
&=&\frac{1}{\gamma L^3}\sum_{\vec{k}}\,
\left\{\frac{e^{\alpha(q^{\ast\,2}-r^2)}}{q^{\ast\,2}-r^2}
+g(\vec{k})\right\}- \frac{1}{\gamma} {\cal
P}\int\frac{d^{\,3}k}{(2\pi)^3}
\left\{\frac{e^{\alpha(q^{\ast\,2}-r^2)}}{q^{\ast\,2}-r^2}
+g(\vec{k})\right\}
\,.
\end{eqnarray}
Here $g$ is a non-singular function of $\vec k$,
which arises from the
$O[(q^{\ast\,2}-k^{\ast\,2})^2]$ terms in eq.\,(\ref{eq:residuerelation}),
with contributions both from the pole and the exponential.
It has an implicit dependence on $q^\ast$.
Its form is not required, however, since the
 sum and integral over $g(\vec k)$ cancel (up to exponentially
small terms in the volume) as a consequence of the Poisson
summation formula. Thus we find that
\begin{equation}
c^{P}(q^{\ast\,2})= \frac{1}{\gamma L^3}\sum_{\vec{k}}\,
\frac{e^{\alpha(q^{\ast\,2}-r^2)}}{q^{\ast\,2}-r^2}-\frac{1}{\gamma}
\,{\cal P}\int\frac{d^{\,3}k}{(2\pi)^3}
\frac{e^{\alpha(q^{\ast\,2}-r^2)}}{q^{\ast\,2}-r^2}\,.
\label{eq:equiv3}
\end{equation}
With this expression for
$c^P(q^{\ast\,2})$, the quantization condition in
eq.\,(\ref{eq:qcfinal}) is manifestly equivalent to that of
ref.\,\cite{rg}.\footnote{%
We use a different regularization, but the result should be
independent of the ultraviolet regulator, and we have checked that
our result agrees numerically with that obtained using the analytic
regularization of ref.~\cite{rg}, up to terms which vanish exponentially
as $L\to\infty$.}

The equivalence can be extended straightforwardly to higher partial waves.
The key point is that the vectors $\vec k^\ast$ and $\vec r$ are equal
at the pole, as established above. Thus one can replace angular
dependence on $\vec k^\ast$ with the same dependence on $\vec r$,
with the difference being non-singular.
The argument goes as follows (assuming $l>0$):
\begin{eqnarray}
c^{P}_{lm}(q^{\ast\,2})&=&\frac{1}{L^3}\sum_{\vec{k}}\,\
\frac{\omega_k^\ast}{\omega_k}\
\frac{e^{\alpha(q^{\ast\,2}-k^{\ast\,2})}}{q^{\ast\,2}-k^{\ast\,2}}
\,k^{\ast\,l}\sqrt{4\pi}\,Y_{lm}(\theta^\ast,\phi^\ast)
\\
&=&\frac{1}{L^3}\sum_{\vec{k}}\,\
\frac{\omega_k^\ast}{\omega_k}\
\frac{e^{\alpha(q^{\ast\,2}-k^{\ast\,2})}}{q^{\ast\,2}-k^{\ast\,2}}
\, k^{\ast\,l}\sqrt{4\pi}\,Y_{lm}(\theta^\ast,\phi^\ast)
\nonumber\\
&&\mbox{}\quad
- \ {\cal P}\int\frac{d^{\,3}k}{(2\pi)^3}\
\frac{\omega_k^\ast}{\omega_k}\
\frac{e^{\alpha(q^{\ast\,2}-k^{\ast\,2})}} {q^{\ast\,2}-k^{\ast\,2}}
\,k^{\ast\,l}\sqrt{4\pi}\,Y_{lm}(\theta^\ast,\phi^\ast)
\\
&=&\frac{1}{\gamma L^3}\sum_{\vec{k}}\,
\left\{\frac{e^{\alpha(q^{\ast\,2}-r^2)}}{q^{\ast\,2}-r^2}
\,r^{\ast\,l}\sqrt{4\pi}\,Y_{lm}(\theta_r,\phi_r)
+g_{lm}(\vec{k})\right\}
\nonumber\\
&&\mbox{}\quad
- \frac{1}{\gamma} {\cal
P}\int\frac{d^{\,3}k}{(2\pi)^3}
\left\{\frac{e^{\alpha(q^{\ast\,2}-r^2)}}{q^{\ast\,2}-r^2}
\,r^{\ast\,l}\sqrt{4\pi}\,Y_{lm}(\theta_r,\phi_r)
+g_{lm}(\vec{k})\right\}
\\
&=&\frac{1}{\gamma L^3}\sum_{\vec{k}}\,
\frac{e^{\alpha(q^{\ast\,2}-r^2)}}{q^{\ast\,2}-r^2}
\,r^{\ast\,l}\sqrt{4\pi}\,Y_{lm}(\theta_r,\phi_r)
\,.
\end{eqnarray}
In the first step we have added an integral that vanishes,
but is nevertheless useful for subsequent manipulations.
In the next step we have rewritten the expression in terms of
$\vec r$ [using spherical components $(r,\theta_r,\phi_r)$],
leading to the presence of the non-singular function $g_{lm}(\vec k)$.
To make the final step,
we note that the sum and integral over $g_{lm}$ cancel,
and that the remaining integral over $\vec k$, which can be rewritten
as an integral over $\vec r$, vanishes by rotational symmetry.
The result is an expression given in ref.\,\cite{rg}, although
regularized differently. The precise relation is
\begin{equation}
c^{P}_{lm}(q^{\ast\,2}) = -
\frac{\sqrt{4\pi}}{\gamma L^3}\left(\frac{2 \pi}{L}\right)^{l-2}
Z^d_{lm}[1; (q^* L/2\pi)^2]
\,, \label{eq:cPvsZd}
\end{equation}
where as usual the equality holds up to terms which
vanish exponentially with the box size.
As already noted in sec.~\ref{subsec:qcgeneral}, with this relation, our general
quantization condition eq.~(\ref{eq:truncateddet}) agrees with
that obtained in ref.~\cite{rg}.

\section{Quantization Condition in Perturbation
Theory}\label{sec:pert}

In this section we test
the validity of the quantization condition in
eq.\,(\ref{eq:qcfinal}) in perturbation theory. We assume
an interaction potential $\lambda \pi^4/4!$ and work to
linear order in $\lambda$.
Specifically, we take the expressions for the s-wave phase-shift
and two pion energy evaluated at $O(\lambda)$ and verify that they
satisfy eq.\,(\ref{eq:qcfinal}).

The $\pi\pi$ phase-shift at
$O(\lambda)$ is readily calculable and is given by
\begin{equation}\label{eq:phaseshift}
\delta(q^\ast)=-\frac{\lambda q^\ast}{16\pi E^\ast}\,.
\end{equation}
We stress that, for this exercise,
we are considering indistiguishable ``pions''
without an isospin degree of freedom.

We now sketch the derivation of the two-pion energy at one-loop
order and then in sec.\,\ref{subsubsec:verification} check that
they satisfy the quantization condition.

\subsection{Evaluation of the Two-Pion Energy in Perturbation
Theory}

For the perturbative evaluation of the energy eigenvalues in a
finite volume, we follow the steps in the derivation
presented in section 4 of ref.\,\cite{spqr}, generalised to an arbitrary
frame.
We refer the reader to ref.\,\cite{spqr} for a more
complete description of the calculation. The
energy eigenvalues are obtained by evaluating the time dependence
of the Euclidean correlation function:\footnote{%
In ref.\,\cite{spqr} the discussion considered the $K\to\pi\pi$ amplitude,
but for the calculation of two pion energies we do not need the initial
kaon.}
\begin{equation}\label{eq:cpertdef}
C_{\vec{q}_1\vec{q}_2}(t)=\langle\,0\,|\,\pi_{-\vec{q}_1}(t)
\pi_{-\vec{q}_2}(t)\,\pi(0)^2\,|\,0\,\rangle\,,
\end{equation}
where
\begin{equation}
\pi_{\vec{q}}(t)=\int\,d^3x\,\pi(\vec{x},t)\,e^{i\vec{q}\cdot\vec{x}}
\end{equation}
is the Fourier transform of the pion field, which creates a pion
with momentum $\vec q$ and destroys a pion with momentum $-\vec q$.
The momenta $\vec q_i$ must take one of the values allowed by
the box size, but are otherwise completely general. In particular,
$\vec P=\vec q_1 + \vec q_2$ need not vanish.

\begin{figure}[t]
\begin{center}
\begin{picture}(300,70)(-10,15)
\Line(30,50)(70,65)\Line(30,50)(70,35)
\GCirc(30,50){1}{0.2}
\GCirc(70,35){1}{0.2}\GCirc(70,65){1}{0.2}
\Text(25,50)[r]{\scriptsize{0}}
\Text(75,65)[l]{\scriptsize{$\pi$}}
\Text(75,35)[l]{\scriptsize{$\pi$}}
\Text(50,25)[t]{\scriptsize{(a)}}
\Oval(210,50)(10,30)(0)
\Line(240,50)(280,65)\Line(240,50)(280,35)
\GCirc(180,50){1}{0.2}
\GCirc(280,35){1}{0.2}\GCirc(280,65){1}{0.2}
\GCirc(240,50){1.5}{0.2}
\Text(175,50)[r]{\scriptsize{0}}
\Text(285,65)[l]{\scriptsize{$\pi$}}
\Text(285,35)[l]{\scriptsize{$\pi$}}
\Text(240,25)[t]{\scriptsize{(b)}}
\Text(210,65)[b]{\scriptsize{$\pi$}}
\Text(210,35)[t]{\scriptsize{$\pi$}}
\end{picture}\end{center}
\caption{Contributions to the correlation function
defined in eq.\,(\ref{eq:cpertdef}):
(a) tree-level diagram; (b) one-loop diagram
giving the energy shift at $O(\lambda)$.}
\label{fig:cpert}\end{figure}
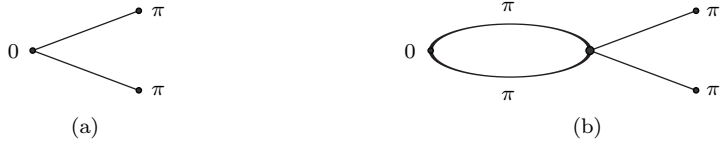

The two-pion energy can be deduced
from the exponential dependence of the
correlation function on $t$. Without loss of generality,
we take $t>0$.
At tree level $C_{\vec{q}_1\vec{q}_2}(t)$ can be
represented by the diagram in fig.\,\ref{fig:cpert}\,(a),
and one obtains
\begin{equation}
C^{(a)}_{\vec{q}_1\vec{q}_2}(t) =
2 \frac{e^{-\omega_1 t}}{2 \omega_1} \frac{e^{-\omega_2 t}}{2 \omega_2}
\,,
\end{equation}
where $\omega_i^2=\vec{q}_i^{\ 2}+m^2$,
and the factor of 2 arises since there are two possible contractions.
The two-pion energy is thus $\omega_1+\omega_2$.

At one-loop order, only the
diagram in fig.\,\ref{fig:cpert}\,(b) gives rise to a shift in the
two-pion energy. Following
ref.\,\cite{spqr}, the contribution linear in $t$ is seen to be
\begin{equation}
C^{(b)}_{\vec{q}_1\vec{q}_2}(t) =
\sum_{{\vec{q}_1}^{\ \prime}\!,\ {\vec{q}_2}^{\ \prime}\atop
{\rm on-shell}}
\,\sum_{t'=0}^{t} \,
-\frac{\lambda}{L^3}
\frac{e^{-\omega'_1 t'}}{2 \omega'_1} \frac{e^{-\omega'_2 t'}}{2 \omega'_2}
 \frac{e^{-\omega_1 (t-t')}}{2 \omega_1} \frac{e^{-\omega_2 (t-t')}}{2 \omega_2}
\,.
\end{equation}
Here ${\vec{q}_1}^{\ \prime}$ and ${\vec{q}_2}^{\ \prime}$ are the loop momenta
(with $\omega'_i$ the corresponding energies), and the sum over
${\vec{q}_i}^{\ \prime}$ is over all ``on-shell'' intermediate states.
These are the choices satisfying
${\vec{q}_1}^{\ \prime}+{\vec{q}_2}^{\ \prime}=\vec P$ and
$\omega'_1+\omega'_2=\omega_1+\omega_2$. The number of such
choices is denoted $\nu$.
In the CM frame, this constraint implies $\omega'_1=\omega'_2=\omega_1=\omega_2$,
so that all contributions are related by rotations, and all
come with the same kinematical factor.
For $\vec P\ne 0$, however, the on-shell constraint can be satisfied in two ways:
(i) those related by a lab frame symmetry transformation, for which
$\omega'_1=\omega_1$ or $\omega'_2=\omega_2$; and (ii)
accidental choices of ${\vec{q}_1}^{\ \prime}$
not related by a lab frame symmetry transformation,
for which $\omega'_1$ equals neither $\omega_1$ or $\omega_2$, although
$\omega'_1+\omega'_2=\omega_1+\omega_2$. The latter are absent in general,
since they require special values of the pion masses.\footnote{%
One example is $\vec{q}_1=(2,4,0)$, $\vec{q}_2=(-2,1,0)$,
${\vec{q}_1}^{\ \prime}=(1,5,0)$ and ${\vec{q}_2}^{\ \prime}=(-1,0,0)$
with $m^2=209/16$ (all quantities in units of $2 \pi/L$).}
They are also absent for the simplest choices of momenta,
e.g. $\vec{q}_1=(1,0,0)$ and $\vec{q}_2=(0,0,0)$ (in units of $2\pi/L$),
for which $\nu=2$ for all pion masses.

In the following, we assume that there are no accidental on-shell configurations.
In this case, the kinematical factor is common to all terms in the sum over $t'$
and one finds that the term linear in $t$ in $C^{(b)}$ is
\begin{equation}
C^{(b)}_{\vec{q}_1\vec{q}_2}(t)=
-t\, \nu\, \frac{\lambda}{L^3}
\frac{e^{-(\omega_1 +\omega_2)t}}{(4 \omega_1\omega_2)^2}
\,.
\end{equation}
Combining this with the result for $C^{(a)}$ one finds, the energy shift
$\Delta E= E-\omega_1-\omega_2$ to be
\begin{equation}\label{eq:deltae}
\Delta E =\frac{1}{L^3}\,\frac{\lambda\nu}{8\omega_1\omega_2}\,.
\end{equation}
Note that this holds for any allowed choice of $\omega_i$.

\subsection{Verification of the Quantization Condition}
\label{subsubsec:verification}

We recall that the quantization condition is
$4\pi \tan(\delta)/q^\ast=1/c^P(q^{\ast\,2})$, since we only
have s-wave scattering.
Verifying this turns out to be simpler if one uses the form of
$c^P(q^{\ast\,2})$ given in eq.\,(\ref{eq:equiv3}), i.e. the
form obtained in ref.\,\cite{rg}. Schematically this form
is $c^P=\gamma^{-1} L^{-3}\sum_{\vec k} (q^{\ast\,2}-r^2)^{-1}$.
At leading order this condition is satisfied almost trivially.
The allowed energies are those of two free pions each with momentum
consistent with the boundary conditions. For any such choice
we have $q^{\ast\,2}=k^{\ast\,2}=r^2$, as discussed in sec.\,
\ref{subsec:comparison}.
Thus $c_P$ is infinite, requiring $\tan\delta=0$,
which is consistent with the leading order result that $\delta=0$.

At next order we consider energies lying close to
one of free pion values: $E=\omega_1+\omega_2+\Delta E$ where
$\omega_i=\sqrt{{\vec q_i}^{\ 2}+m^2}$, with $\vec q_i$ being
the allowed finite-volume momenta. The energy shift $\Delta E$ is of $O(\lambda)$.
Thus in each of the $\nu$ terms in the summation which
at leading order had $q^{\ast\,2}=r^2$ the denominators
$q^{\ast\,2}-r^2$ are now of $O(\lambda)$, and hence
$c_P$ is of $O(1/\lambda)$.
All other terms in the summation, as well
as the principal part integral required for regularization, are of $O(1)$
and thus subleading.
Keeping only the leading terms, the quantization condition becomes
\begin{equation}
\delta = \frac{q^\ast}{4\pi}\, \frac{\gamma L^3}{\nu}\,
\Delta E \,\frac{d(q^{\ast\,2}-r^2)}{dE}\bigg|_{q^{\ast\,2}=r^2}
\,.
\end{equation}
Here we have made use of the result, to be established below, that
all $\nu$ relevant terms give the same contribution to $c^P$.

The first part of the derivative is readily evaluated.
Recalling that $4(q^{\ast\,2}+m^2)=E^{\ast\,2}= E^2-P^2$,
and noting that the derivative is to be performed at fixed $P$ and $m$,
we have
\begin{equation}
\frac{d q^{\ast\,2}}{dE} =  \frac{E}{2} \,.
\label{eq:derivq}
\end{equation}
Next, using the definition (\ref{eq:rdef}) for $\vec r$, and noting
the $\vec k$ is held fixed, we have
\begin{equation}
\frac{d r^2}{dE}=\frac{d r_\parallel^2}{dE}
=(k_\parallel - P/2)^2 \frac{d \gamma^{-2}}{dE}
= (k_\parallel - P/2)^2 \frac{2 P^2}{E^3}
\,.
\label{eq:derivr}
\end{equation}
Combining (\ref{eq:derivq}) and (\ref{eq:derivr}) we find
\begin{equation}
\frac{d(q^{\ast\,2}-r^2)}{dE} = \frac{E^4 - (2 k_\parallel - P)^2 P^2}{2E^3}
\,.
\end{equation}
This expression simplifies when
evaluated for two on-shell pions (i.e. for $q^{\ast\,2}=r^2$).
The four-vectors of the two pions in the lab frame are then
\begin{equation}
q_1^\mu=(\omega_1,\vec{k})\,,\quad
q_2^\mu=(\omega_2,\vec{P}-\vec{k})
\,.
\end{equation}
Observing that
\begin{equation}
0 = (q_1+q_2)^\mu (q_1-q_2)_\mu = E(\omega_1-\omega_2) -
(2 k_\parallel-P)P
\,,
\end{equation}
the derivative at the pole is readily seen to be
\begin{equation}
\frac{d(q^{\ast\,2}-r^2)}{dE}\bigg|_{q^{\ast\,2}=r^2}
=\frac{E^2-(\omega_1-\omega_2)^2}{2E}
=\frac{2\omega_1\omega_2}{E}
\,.
\end{equation}
Note that this result is the same for all $\nu$ contributions
to the sum in $c^P$.

Inserting this result in the quantization condition and rearranging
gives
\begin{equation}
\delta = \frac{q^\ast}{16\pi E^\ast}\, \frac{\Delta E L^3}{\nu}\,
8\omega_1\omega_2
\,.
\end{equation}
This is consistent with the results for $\delta$ and $\Delta E$
given respectively in eqs.\,(\ref{eq:phaseshift}) and (\ref{eq:deltae}) above,
confirming the validity of the quantization condition
at $O(\lambda)$ in perturbation theory.

\section{Finite-Volume Effects in Matrix Elements}
\label{sec:ll}

In this section we generalise the Lellouch-L\"uscher
formula~\cite{ll} for the finite-volume effects in $K\to\pi\pi$
matrix elements from the centre-of-mass frame to a moving frame in
which the two-pions have total momentum $\vec{P}$.
We find the following relation between the matrix elements of the effective
Weak Hamiltonian density, ${\cal H}_W$, in infinite and finite volumes:
\begin{equation}
|A_W|^2=
8 \pi V^2\,\frac{m_KE^2}{q^{\ast\,2}}\left\{\delta^\prime(q^\ast)+
\phi^{P\,\prime}(q^\ast)\right\}\ |M_W|^2
\,, \label{eq:llmoving}
\end{equation}
where the $^\prime$ represents
the derivative with respect to $q^\ast$,
$\delta$ is the s-wave phase shift,
$\phi^P$ the kinematic function defined in eq.~(\ref{eq:qcfinal}),
$V=L^3$ is the spatial volume, and
the $K\to\pi\pi$ matrix elements in infinite and finite
volumes are respectively
\begin{equation}
A_W = \,_\infty\langle\,\pi\pi;E,\vec{P}\,|
\,{\cal H}_W(0)\,|\,K;\vec{P}\,\rangle_\infty
\label{eq:adef}
\end{equation}
and
\begin{equation}
M_W = \,_V\langle\,\pi\pi;E,\vec{P}\,|
\,{\cal H}_W(0)\,|\,K;\vec{P}\,\rangle_V
\label{eq:mdef}
\end{equation}
(as indicated by the subscripts).
The infinite volume states have the usual relativistic normalizations,
while the finite volume states are normalized to unity.
Note that for this formula we are assuming that ${\cal H}_W(0)$ does not
insert energy or momentum. In particular, for both infinite and finite
volume the energy of the two-pion state equals that of
the initial kaon, $E^2= m_K^2 + P^2$.
We comment on the generalisation below.

The result (\ref{eq:llmoving}) is valid, as usual, up to corrections
vanishing exponentially with $L$. It also assumes that one can neglect
the scattering amplitude in the $l=2,4,\dots$ waves at $E^\ast=m_K$.
Note that this is a stronger assumption for $\vec P\ne 0$ than for $\vec P=0$.
In the latter case, the first higher partial wave which contributes has $l=4$,
due to the cubic symmetry. For a moving frame, however, the symmetry
is reduced and the leading contribution comes from $l=2$.
Technically this arises because $c^P_{20}\propto Z^d_{20}\ne 0$, as shown
explicitly in ref.~\cite{rg}.

In the centre-of-mass frame the original derivation of the formula
in eq.\,(\ref{eq:llmoving}) proceeded by using degenerate
perturbation theory\,\cite{ll}, in which the volume was chosen so
that, in the absence of weak interactions, the kaon and the
two-pion states are degenerate. This degeneracy is then broken by
the weak interactions. In the derivation of ref.\,\cite{lmst} the
factor relating the infinite and finite volume matrix elements was
interpreted in terms of the density of states of the two-pion
system. We now derive eq.\,(\ref{eq:llmoving}) using both
approaches in turn. For simplicity, we continue to treat the pions
as identical, ignoring the isospin degree of freedom. The final
result applies, however, separately for each isospin state (using
the corresponding phase shift).

\subsection{Derivation
using degenerate perturbation theory}

In this section we generalise the derivation of ref.\,\cite{ll}
for the finite-volume effects in matrix elements to a moving
frame. We start by considering the case with no weak interactions
(so that $K\to\pi\pi$ decays are not possible) and take the
volume such that the energy of a kaon with momentum $\vec P$
is equal to the energy of one of
the finite-volume two-pion states with the same total momentum.\footnote{%
We are considering here only the two-pion states with an s-wave
component, so that their energy is determined by the quantization
condition eq.\,(\ref{eq:qcfinal}) using the s-wave scattering
amplitude. These states go over to pure s-wave states as the
volume is sent to infinity.}
The weak interactions are now introduced as an
(arbitrarily weak) perturbation which
breaks the degeneracy between the kaon and two-pion states. The
energy eigenvalues are then
\begin{equation}
E=E_0\pm \lambda V|M_W|
\,, \label{eq:epm}
\end{equation}
where
$E_0^2=P^2+m_K^2$ and $\lambda$ is a parameter introduced to
facilitate the counting of orders of perturbation theory and is
set to 1 in the final results. The factor of $V$ is present
because in eq.\,(\ref{eq:mdef}) we have defined $M_W$ to be the
matrix element of the Hamiltonian density.

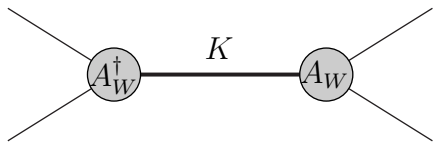
\begin{figure}[t]
\begin{center}
\begin{picture}(200,60)(0,25)
\SetWidth{1.5}\Line(60,50)(140,50) \SetWidth{0.5}
\Line(20,75)(60,50)\Line(20,25)(60,50)\Line(140,50)(180,75)
\Line(140,50)(180,25) \GCirc(60,50){10}{0.8}
\GCirc(140,50){10}{0.8}\Text(60,50)[]{$A_W^\dagger$}
\Text(140,50)[]{$A_W$}\Text(100,56)[b]{$K$}
\end{picture}
\caption{$O(\lambda)$ contribution to the $\pi\pi$ scattering
amplitude (see text). The thick line represents the propagator of
a kaon, which is off-shell by an amount of $O(\lambda)$. $A_W$
represents the infinite-volume $K\to\pi\pi$ matrix
element.\label{fig:pipiscat}}
\end{center}
\end{figure}

We can also calculate the shifts in (\ref{eq:epm}) by taking into account
the effect of the weak interactions on the phase shift, which then
impacts the finite-volume energies through the quantization condition.
The only contribution to the scattering amplitude
(and thus phase shift) linear in $\lambda$ is that
from the diagram shown in fig.\,\ref{fig:pipiscat}\,\cite{ll}. The
energy and momentum of the external $\pi\pi$ states are chosen to be
$(E,\vec{P})$, where $E$ is one of the finite volume energies in
eq.\,(\ref{eq:epm}). The kaon propagator is thus off shell by $O(\lambda)$,
and when combined with the two $K\pi\pi$ vertices, each of $O(\lambda)$, the
result is a contribution to the s-wave scattering amplitude
of first order in perturbation theory:
\begin{equation}
\Delta M_s = \mp\lambda\frac{|A_W|^2}{2E_0V|M_W|}e^{2i\delta(q^\ast)}\,,
\end{equation}
where $\delta$ represents the unperturbed phase-shift. Using
eq.\,(\ref{eq:mtodelta}) we obtain the expression for
$\bar\delta$, the perturbed phase-shift (i.e. the phase-shift
including $O(\lambda)$ terms):
\begin{equation}
\bar{\delta}(q^\ast)=\delta(q^\ast)
\mp\frac{\lambda}{32\pi}\,\frac{q^\ast|A_W|^2}{m_KE_0V|M_W|}\,.
\label{eq:deltapert2}\end{equation}
In order to obtain eq.\,(\ref{eq:llmoving}) we
\begin{enumerate}
\item use the quantization condition to replace
$\bar{\delta}(q^\ast)$ by $n\pi-\phi^P(q^\ast)$; \item expand
$\phi^P(q^\ast)$ and $\delta(q^\ast)$ around the unperturbed point
$q_0^\ast$, where $E_0^{\ast\,2}=m_K^2=4q_0^{\ast\,2}+m^2$,
\begin{equation}
\phi^P(q^\ast)\simeq\phi^P(q_0^\ast)+\Delta
q^\ast\,\phi^{P\,\prime}(q_0^\ast)\quad\textrm{and}\quad
\delta(q^\ast)\simeq\delta(q_0^\ast)+\Delta
q^\ast\,\delta^{\prime}(q_0^\ast)
\end{equation}
with \begin{equation} \Delta q^\ast=\pm\frac{\lambda
E}{4q^\ast}\,V|M_W|\,;
\end{equation}
\item Note that $\phi^P(q_0^\ast)=n\pi-\delta(q_0^\ast)$ and
equate the $O(\lambda)$ terms on both sides of
eq.\,(\ref{eq:deltapert2}).
\end{enumerate}

\subsection{Derivation %
of eq.\,(\ref{eq:llmoving})
using the density of states}

In this section we derive eq.\,(\ref{eq:llmoving}) by generalising
the arguments of ref.\,\cite{lmst}. Let $\sigma(\vec{x},t)$ be an
interpolating operator which can create two-pion states from the
vacuum and consider the two point correlation function in
finite volume:
\begin{eqnarray}
C_{\sigma}^V(t)&\equiv&\int d^3x\,e^{ i\vec{P}\cdot\vec{x}}\,
\langle\,0\,|\,\sigma(\vec{x},t)\,\sigma(\vec{0},0)\,|\,0\,\rangle_V\\
&=&V\sum_n\,|\,\langle\,0\,|\,\sigma(0)\,|\,\pi\pi;\vec{P},n\,\rangle\,|^{\,2}\,
e^{-E_nt}\\
&\mathrel{\mathop{\kern0pt\longrightarrow}\limits_{V\to \infty}}&
V\int\,dE\,\rho_V(E)\,|\,\langle\,0\,|\,\sigma(\vec{0},0)\,|\,
\pi\pi;E,\,\vec{P}\rangle_V\,|^{\,2}\,e^{-Et}\,.\label{eq:csigmados}
\end{eqnarray}
In eq.\,(\ref{eq:csigmados}) $\rho_V(E)$ is the density of
states and can be evaluated using the quantization condition in
eq.\,(\ref{eq:qcfinal}). This condition can be rewritten as
\begin{equation}
\delta(q^\ast)+\phi^P(q^\ast)= n \pi\,,
\end{equation}
with $n$ an integer. We argue below that, for large
enough $L$, $\delta+\phi^P$ increases monotonically,
so that $n$ counts the number of states.
Given this, the density of states is given by
\begin{equation}
\rho_V(E)=\frac{dn}{dE}=\frac{E}{4\pi q^\ast}
\left\{\delta^\prime(q^\ast)+\phi^{P\,\prime}(q^\ast)\right\}
\,.
\end{equation}

The monotonicity of $\delta+\phi^P$ follows because,
as explained below,
$\phi^P$ is a monotonically increasing function of $q^\ast$
whose derivative is proportional to $L$. By contrast the dependence
of $\delta$ on $q^\ast$ is volume independent.
Thus, although $\delta$ need not be an increasing function of $q^\ast$,
its dependence will be overwhelmed by that of $\phi^P$ for
large enough $L$.

The behaviour of $\phi^P$ can be seen most easily using
the form obtained in ref.~\cite{rg}:
\begin{equation}
\phi^P(q^\ast) = \tan^{-1}\left[
q^\ast/(4\pi c^P)\right]\,,\qquad
c^P = (\gamma L^3)^{-1} \sum_{\vec k} (q^{\ast\,2}-r^2)^{-1}
\,.
\end{equation}
Aside from the factor of $\gamma^{-1}$, $c^P$ decreases monotonically
between the poles which occur whenever there is a free two-pion
state in volume $L^3$ with total momentum $\vec P$.
As $L$ increases, the spacing in $q^\ast$ between such states decreases,
so the derivative $c^{P\,\prime}$ becomes more negative,
and overwhelms any positive contribution to the derivative
from $\gamma^{-1}$. This behaviour of $c^P$ implies that
$\phi^P$ increases monotonically at large enough $q^\star$.
Furthermore, since $\phi^P$ increases by $\pi$ for each
transit between poles, whose spacing $\sim 1/L$,
the derivative $\phi^{P\,\prime}$ is proportional to $L$.

We now return to the main line of the argument.
The infinite-volume correlation function corresponding
to $C_\sigma^V$ is given by
\begin{equation}
C_\sigma^\infty(t)=\frac{1}{16\pi^2}\int\,dE\,\frac{q^\ast}{E^\ast}\,
|\,\langle\,0\,|\,\sigma(\vec{0},0)\,|\,
\pi\pi;\vec{P},E\,\rangle_\infty\,|^{\,2}\,e^{-Et}\,,\label{eq:csigmados2}
\end{equation}
where $E^{\ast\,2}=4q^{\ast\,2}+m^2=E^2-P^2$ as above.
Comparing eqs.\,(\ref{eq:csigmados}) and (\ref{eq:csigmados2}) we
obtain the correspondence between the infinite-volume and
finite-volume kets:
\begin{equation}
|\, \pi\pi;\vec{P},E\,\rangle_\infty \Leftrightarrow
4\pi\sqrt{\frac{VE^\ast \rho_V(q^\ast)}{q^\ast}}
|\,\pi\pi;\vec{P},E\,\rangle_V\,.\label{eq:double}
\end{equation}

Similarly, by considering correlation functions of interpolating
operators for single particle states, we find:
\begin{equation}
|K;\vec{P}\,\rangle_\infty\Leftrightarrow\sqrt{2EV}
|K;\vec{P}\,\rangle_V\,,\label{eq:single}
\end{equation}
with $E^2=P^2+m_K^2$. Eq.\,(\ref{eq:single}) is simply the
relation between the wavefunctions of states normalized
relativistically ($2E$ particles per unit volume) and
non-relativistically in a finite-volume (1 particle in volume
$V$).

Combining eqs.\,(\ref{eq:double}) and (\ref{eq:single}) we obtain
the required relation between the matrix elements in infinite and
finite-volume in eq.(\ref{eq:llmoving})\,.

In eq.\,(\ref{eq:llmoving}) the matrix elements correspond to
physical processes and energy and momentum are conserved. However,
eqs. (\ref{eq:double}) and (\ref{eq:single}) can be applied more
generally. One such application is to the chiral extrapolation of
the matrix elements computed in lattice simulations. Since these
are currently performed at unphysically large masses for the $u$
and $d$ quarks, the results have to be extrapolated to the
physical point. Chiral perturbation theory ($\chi$PT) can be
useful in guiding this extrapolation. For example, in
ref.\,\cite{spqr} it was proposed to fix the low-energy constants
which appear in $\chi$PT at next-to-leading order by computing
$K\to\pi\pi$ matrix elements with the kaon and one of the final
state pions at rest and varying the momentum of the second pion.
Equations (\ref{eq:double}) and (\ref{eq:single}), with different
energies and momenta for the kaon and two-pion system, can be
combined to determine the finite-volume effects also in such
cases.

\section{Summary and Conclusions}
\label{sec:conclusions}

In this paper, we have provided a field theoretic derivation of the finite
volume energy shift for two hadron states in a moving frame, confirming the result
obtained by ref.~\cite{rg} using a relativistic quantum mechanical approach.
We have also determined the finite-volume corrections in matrix elements of local
composite operators with an initial and/or final state consisting of
two hadrons, thus generalising the Lellouch-L\"uscher factor to moving frames.
The path is therefore now open to numerical studies of two pion energies
and matrix elements in a moving frame. As described in the introduction,
there are a number of important applications which depend upon, or are greatly
simplified by, working with $\vec P\ne0$ as well as  $\vec P=0$.

Our work provides three new results.
The first is technical: we give
a simple derivation of the summation formulae needed to determine
finite-volume effects. It has the advantages of
conceptual simplicity, and that it can be readily applied in any frame.
In addition, it provides a
straightforward way of calculating the required regularized sums numerically.
It is of comparable numerical efficiency to the method used in
refs.~\cite{ml3} and \cite{rg}, based on analytic regularization.

Our second new result is a method to determine the energies of
finite volume states, to all orders in perturbation theory in
$1/L$, based entirely in field theory. This is in contrast to the
method developed in refs.~\cite{ml2,ml3} for the centre-of-mass
frame. There it is shown that, in the centre-of-mass frame, the
two-particle energies in a general relativistic field theory are
related to those of an auxiliary
 non-relativistic quantum mechanical (NRQM) system in the
same finite volume, with an energy-dependent potential chosen to
reproduce the relativistic scattering phase shifts. The analysis
is then carried out in the NRQM system. Perturbation theory in
$1/L$ is developed in ref.~\cite{ml2}, with ref.~\cite{ml3}
providing the generalisation to all orders in $1/L$. The use of a
NRQM theory is a technical device, and does not require the
underlying two-particle system to be non-relativistic. It reflects
the fact that the dominant finite volume effects come from nearly
on-shell two-particle intermediate states. Nevertheless, we
believe that it is conceptually simpler to avoid the need for an
auxiliary NRQM theory, and this is what our approach allows. In
addition, we obtain the general result with multiple non-vanishing
partial wave amplitudes in a very straightforward fashion.

The particular advantage of our approach, however, is that it is readily
generalisable to a moving frame. The general formalism remains
unchanged, but the required kinematical functions,
which arise from the relation between
two-particle loop momentum sums and integrals, depend on $\vec P$.
The generalisation of the method
using the auxiliary NRQM theory is less obvious,
in part because of the mismatch
between relativistic and non-relativistic energies for $\vec P\ne0$.
This is finessed in ref.~\cite{rg} by the use of a {\em relativistic} two particle
wave function, satisfying the Klein-Gordon equation. The wavefunction
 is subject to periodic
boundary conditions in the moving frame, and the key step is to convert these
to boundary conditions in centre-of-mass coordinates.
One would expect the latter to involve unequal times because of
the Lorentz boost, but the authors of ref.~\cite{rg} argue that this is
not the case. Their argument is based on the observation that
outside of the interaction region the wave function is that of two
free particles and can readily be shown to be independent of the
relative time in the centre-of-mass frame. From this they conclude
that the boundary conditions can be applied at equal times also in
the centre-of-mass frame, with the effect of the boost causing an
elongation of the box in the direction of $\vec{P}$. We were left
with the question as to whether this procedure correctly
corresponds to periodic boundary conditions in the moving frame
(including inside the interaction region) and this question
partially motivated us to develop the alternative approach
presented here. Since our final results for the quantization
condition can be brought into the same form as those of ref.~\cite{rg},
we have indirectly confirmed their assumptions.

As a check of our result (and thus also that of ref.~\cite{rg}) we
have calculated the two-pion energies to leading non-trivial order
in the perturbative expansion in a $\lambda\phi^4$ theory, and
confirmed the quantization formula in a general moving frame.

Our final new result is the generalization to moving frames of the
Lellouch-L\"uscher factor which contains the finite-volume effects
in $K\to\pi\pi$ matrix elements.
This turns out to require only a simple kinematical change: the
non-trivial effects of working in a moving frame are already incorporated
in the functions $c^P_{lm}$ that determine the two-particle energies
in finite volumes.

\subsection*{Acknowledgements}
We thank Norman Christ and Takeshi Yamazaki for communicating their
work to us prior to its release. We thank them, and also
Maarten Golterman, Kari Rummukainen and Massimo Testa
for helpful discussions and correspondence.
CTS thanks the Institute for Nuclear Theory for its hospitality
during the Programme on \textit{Effective Field Theories, QCD and
Heavy Hadrons} and the Department of Energy for partial support
during the later stages of this work.
SRS thanks the University of Southampton for hospitality during the
initial part of this work and acknowledges support from a
PPARC Visiting Fellowship (PPA/V/S/2003/00006).
We also acknowledge support from PPARC grants PPA/G/O/2002/00468
and PPA/G/S/2003/00093 and from DOE grant DE-FG02-96ER40956.


\end{document}